\documentclass[a4paper,11pt]{article}
\pdfoutput=1 

\usepackage{jinstpub} 

\graphicspath{{./figures/}}

\title{\boldmath Machine-learning-based prediction of parameters of secondaries in hadronic showers using calorimetric observables}

\author[a,1]{M. Chadeeva,\note{Corresponding author.}}
\author[a,b]{S. Korpachev}

\affiliation[a]{P.N. Lebedev Physical Institute of the RAS,\\53, Leninskyi prospekt, Moscow, Russia}
\affiliation[b]{Moscow Institute of Physics and Technology,\\Dolgoprudnyi, Russia}

\emailAdd{chadeevamv@lebedev.ru}

\abstract{The paper describes a novel neural-network-based approach to study the distributions of secondaries produced in hadronic showers using observables provided by highly granular calorimeters. The response is analysed of the highly granular scintillator-steel hadron calorimeter to negative pions with momenta from 10 to 80 GeV simulated with two physics lists from the Geant4 package version 10.3. Several global observables, which characterise different aspects of hadronic shower development, are used as inputs for a deep neural network. The network regression model is trained using a supervised learning and exploiting true information from the simulations. The trained model is applied to predict a number of neutrons and energy of neutral pions produced within a hadronic shower. The achieved performance and possible application of the model to validation of simulations are discussed.}

\keywords{Calorimeter methods, Detector modelling and simulations I}

\arxivnumber{2205.12534} 

\begin{document}
\maketitle
\flushbottom

\section{Introduction}
\label{sec:intro}

The simulation is a very important stage in the development of calorimeters and other detector subsystems for the experiments in particle physics. The main goal of simulations is to optimise a layout to meet the expected performance of the detector. The important stage, which follows a physical process generation, is the particle propagation through detector materials and producing the signals measured by active detector elements. The widely used tool for simulation of particle propagation through matter is the Geant4 package~\cite{ref:Geant4-2003}. This package is known to provide a very precise simulation of ionisation losses as well as electromagnetic shower development, which properties are predicted with percent-level accuracy~\cite{ref:Geant4-2016}. The simulation of hadronic showers produced by the inelastic interactions of hadrons with nuclei is much more complicated and involves the complex combination of theoretical and phenomenological models. The latest versions of Geant4 enable a very good modelling of mean values for the main calorimetric observables such as reconstructed energy. At the same time, the development and implementation of the Particle Flow (PF) approach~\cite{ref:PFA-2009ilc} for the jet energy reconstruction in the modern detector systems with highly granular calorimeters poses new requirements to the simulation precision. The PF algorithms fully exploit the information from the most precise systems as trackers and minimise the impact from hadron calorimeter as the less precise subdetector. But to leverage of the PF jet energy reconstruction, one needs to disentangle contributions from different particles in the calorimeters, which is based on both high calorimeter granularity and good understanding of the radial and longitudinal shower development. Therefore, correct modelling of hadronic shower substructure is relevant for the estimate and prediction of the PF algorithm performance.

The validations of simulations are performed using experiments on test beams of particles with known initial momenta. The highly granular electromagnetic and hadron calorimeters developed by the CALICE collaboration  \cite{ref:SiECAL-2008,ref:AHCAL-2010cc,ref:SDHCAL-2015,ref:ScECAL-2018} are a new generation of instruments, which allow visualisation of hadronic shower substructure and  validation of simulations in more details than just reconstructed energy and resolution. The high longitudinal and radial segmentation of such devices helps to identify track segments within a shower~\cite{ref:AHCAL-2013tracks,ref:SDHCAL-2017tracks}, to analyse the shower energy density distribution and find the position of the first inelastic interaction in the calorimeter~\cite{ref:AHCAL-2013valid,ref:SiWECAL-2015,ref:AHCAL-2016dec,ref:WAHCAL-2015,ref:DHCAL-2019}. The previous studies of showers initiated by hadrons with the energies from several GeV to more than hundred GeV in highly granular calorimeters and comparisons with Geant4 simulations lead to the conclusion that the differences between data and simulations increase with energy and simulations predict narrower hadronic showers than observed in experimental data.

Given the diversity of existing hadronic models, there are several parameters of hadronic showers, which cannot be directly reconstructed in real data, but are of interest for comparisons between data and simulations. For example, the number of generated secondary neutrons within a shower is related to the shower topology and time structure, while the energy of neutral pions defines the amount of electromagnetic fraction in a hadronic shower. 
A comparison of these parameters can help in better understanding of agreement between data and a particular hadronic model in terms of the properties of secondary particles. 
As the relationships between parameters of interest and observables are quite complicated and nonlinear, the natural choice for this task is a machine learning (ML) approach. The ML algorithms are widely used to solve classification and regression problems in high energy physics \cite{ref:ML-2002,ref:ML-2018review,ref:ML-2021gnn}. For the calorimetry, the focus is on fast simulations and jet energy reconstruction \cite{ref:ML-2020calo,ref:ML-2020bjet,ref:ML-2022kinem}. In this paper, the relationships are exploited between observables in highly granular hadron calorimeter and properties of secondary particles produced within a hadronic shower. The calorimetric observables are then used as inputs to regression model implemented in the deep neural network (DNN) architecture. The model is trained on simulated hadronic showers using a supervised learning to predict the distributions of number of neutrons and energy of neutral pions. The model performance achieved is demonstrated on simulated 40~GeV pion showers.

The paper is organised as follows. The section \ref{sec:model} describes simulation details, including calorimeter model and digitisation, Geant4 hadronic models and extraction of particle properties at generation level. New calorimetric observables, which can be studied in highly granular calorimeters, are defined in section \ref{sec:calo_obs}. The relationship between calorimetric observables and secondary particle properties is demonstrated in section \ref{sec:corr} and the regression model implemented in the deep neural network to make use of these relationships is presented in section \ref{sec:dnn}. The section \ref{sec:res} shows the results of DNN training and predictions of properties of secondaries in a hadronic shower using calorimetric observables.

\section{Calorimeter model and simulation conditions}
\label{sec:model}

Hadronic showers induced by single negative pions with energies from 10 to 80~GeV were simulated in the model of the highly granular tile hadron calorimeter. The model reproduces the real CALICE scintillator-steel analog hadron calorimeter prototype (AHCAL), which belongs to the second generation of such devices successfully developed and tested by the CALICE collaboration~\cite{ref:procVCI-2019}. The design of the calorimeter is briefly described below and is precisely modelled in the simulations. 

\subsection{Highly granular analog hadron calorimeter}
\label{sec:calo_ahcal}

The highly granular calorimeter AHCAL consists of 39 active planes assembled from small scintillator tiles 30$\times$30$\times$3~mm$^3$. Each tile has a dimple (spherical, 7~mm in diameter and 1.6~mm deep) in one of the big surfaces and is wrapped in the reflective foil. The scintillator light from the tile is read out by a silicon photomultiplier placed in front of the dimple. The active planes are interleaved with 2-cm-thick steel plates. The transverse size of the prototype is 72$\times$72 cm$^2$ (576 tiles per plane, 22188 channels in total). The longitudinal depth of the standalone device amounts up to 4.3$\lambda_{\mathrm{I}}$ (where $\lambda_{\mathrm{I}}$ is the nuclear interaction length). The AHCAL was commissioned in 2018 and tested with electron, muon and pion test beams at CERN SPS \cite{ref:AHCALtest}. The equalisation of cell response in highly granular calorimeters is performed by exposing the calorimeter to minimum ionising particles (MIP). For cell response calibration, the most probable value is estimated of the distribution of signal produced by muons in each cell of the experimental setup. The details of this calibration procedure can be found in ref.~\cite{ref:AHCAL-2010cc}. 

\subsection{Digitisation and event preselection}
\label{sec:conditions}

The simulation packages provide the response in a particular active cell in units of GeV. The most probable value of this simulated response to muons is taken as a conversion factor from GeV to MIP for simulations. To account for detector effects, such as light collection efficiency, photon detection efficiency, pixel statistics and electronic noise, additional channel-wise smearing is applied in the dedicated digitisation module of the simulation software. The level of smearing is tuned using comparisons of simulated and experimental response to muons. Finally, the output of simulated cell amplitudes is given in units of MIP as in data. Only cells with signals above the 0.5~MIP threshold are used for the further analysis.   
 
The primary $\pi^{-}$ particles strike the calorimeter perpendicular to and near the centre of the front face. 
To study the shower substructure, the minimisation of both longitudinal and transverse leakage from the calorimeter is very desirable. The transverse leakage is negligible due to large transverse size of the detector and beam position near the center of the front face. There are two options for minimisation of the longitudinal leakage. The first one is to extend the depth of the model by increasing the number of layers compared to the real detector. The second option is to select events where hadronic shower starts developing not far from the beginning of the calorimeter as the fine granularity of the AHCAL allows identification of the position of the first inelastic interaction with appropriate precision~\cite{ref:AHCAL-2016dec}. The second option is preferred as it allows further comparison with experimental data when available. In this study, the events are preselected for the analysis with the identified first inelastic interaction (shower start) in the layers from 3 to 6 from the calorimeter front. Such a choice is a compromise between the minimisation of longitudinal leakage and the reliability of shower start identification, which is less accurate in the first two layers due to the lack of information about a primary particle track.

\subsection{Geant4 hadronic models}
\label{sec:physlist}

The Geant4 package contains a number of models, which describe the propagation of different particles through detector material and their interactions with matter.  No model allows appropriate description for all species in a wide energy range and, therefore, models are grouped in the so-called physics lists, which change from version to version. In this study, the simulations of hadronic showers in the AHCAL were performed using the Geant4 package version 10.3. Two physics lists are considered for comparisons: FTFP\_BERT\_HP and QGSP\_BERT\_HP, where FTF states for the Fritiof parton model, BERT for the  Bertini intranuclear cascade and QGS for the quark-gluon string model. The letter "P" corresponds to the Precompound model, which is invoked to de-excite the remnant nucleus after the initial high energy interaction. The suffix HP indicates that High Precision neutron models and cross sections are used to describe elastic and inelastic scattering, capture and fission processes for neutrons of 20 MeV and below.

The FTFP\_BERT physics list is recommended for collider physics applications and is the default option in Geant4 since 2016. The Bertini intranuclear cascade is responsible for hadron interactions between 0 to 6 GeV/nucleon, while  the FTF model works over the range from 3 to 100 TeV/nucleon. In the overlapping region, the Bertini model is invoked with a probability that decreases linearly from 1.0 to 0.0 with increasing energy as described in more detail in ref.~\cite{ref:G4-2016ftfp}.
The QGSP\_BERT physics list is the former default in Geant4. It implements the BERT model for hadron interactions between 0 to 6 GeV, QGS model for protons, neutrons, pions and kaons above 12 GeV.  The FTF model handles these same particles over the range from 3 GeV to 25 GeV and also anti-particles in the range from 0 to 100 TeV/nucleon~\cite{ref:G4-2009qgsp}. The overlapping energy regions between FTF and BERT models as well as between QGS and FTF models are treated as described above for FTFP\_BERT physics list. 

\subsection{Properties of secondary particles in hadronic showers}
\label{sec:secondaries}

\begin{figure}[htbp]
\centering 
\includegraphics[width=.45\textwidth]{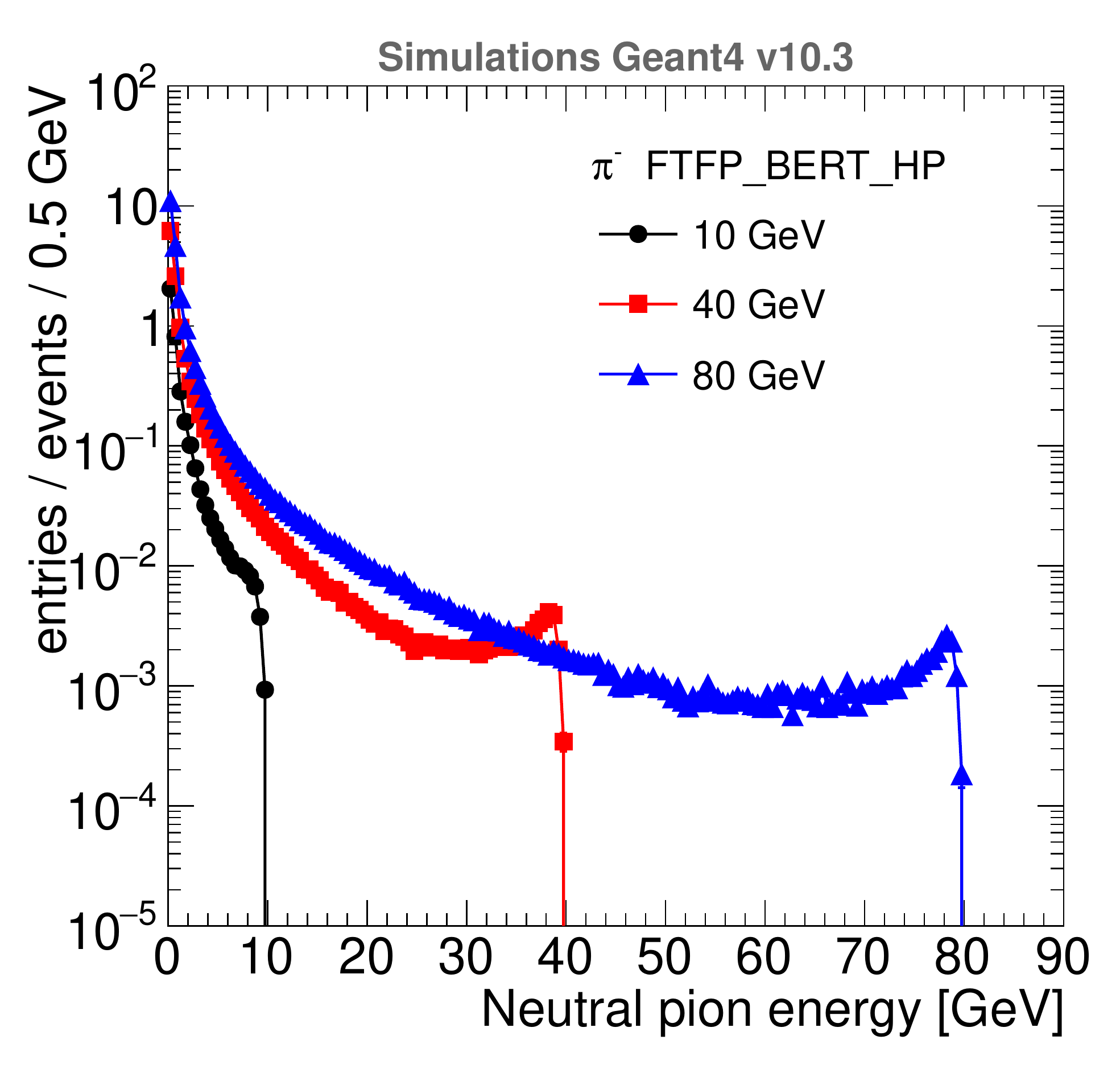}
\qquad
\includegraphics[width=.45\textwidth]{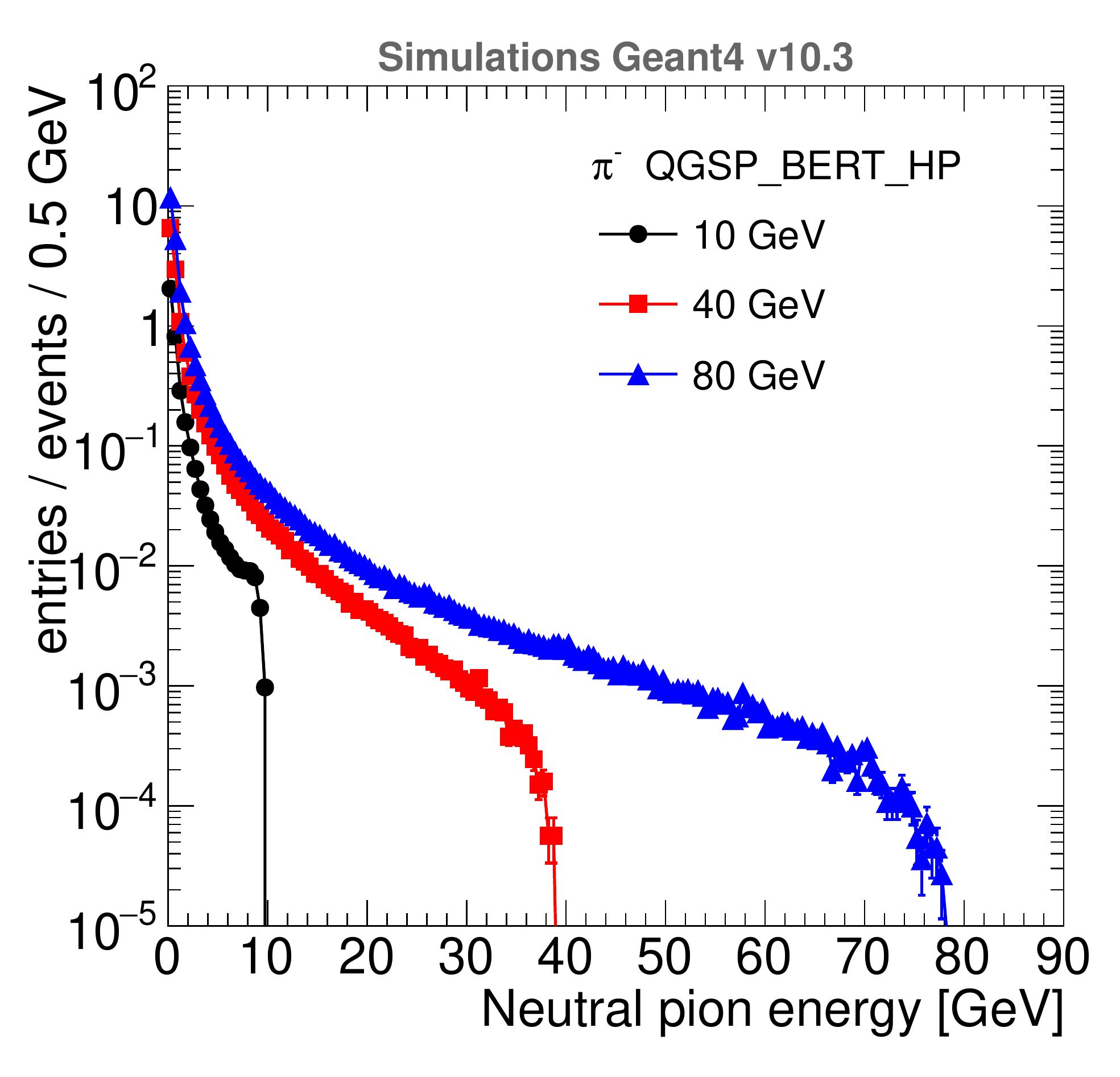}
\caption{\label{fig:Epi0} Energy spectra of neutral pions produced in a hadronic shower initiated by $\pi^{-}$ with energy of 10~GeV (black), 40~GeV (red) and 80~GeV (blue) as simulated using FTFP\_BERT\_HP (left) or QGSP\_BERT\_HP (right) physics lists of Geant4 version 10.3.}
\end{figure}

\begin{figure}[htbp]
\centering 
\includegraphics[width=.3\textwidth]{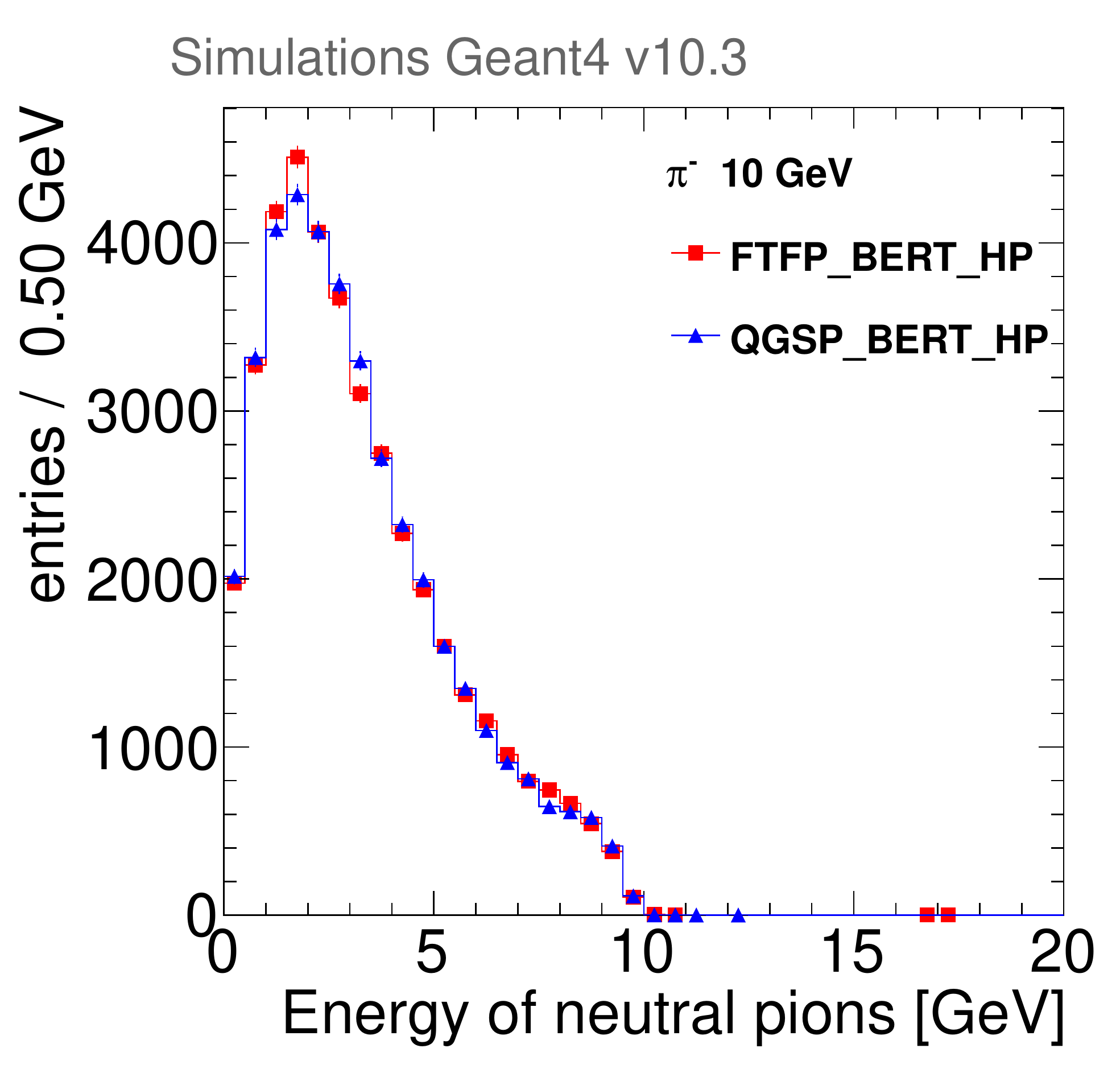}
\includegraphics[width=.3\textwidth]{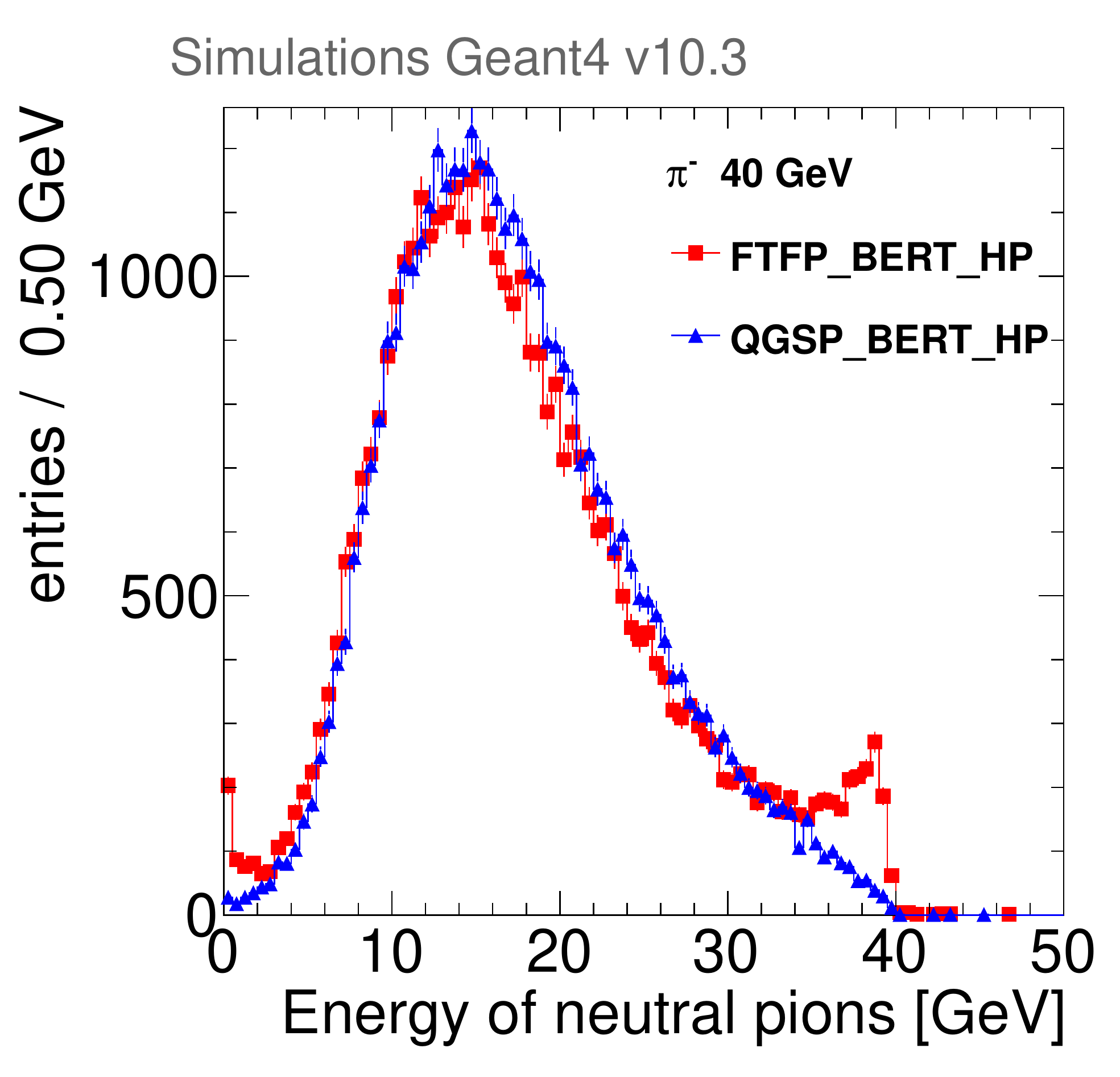}
\includegraphics[width=.3\textwidth]{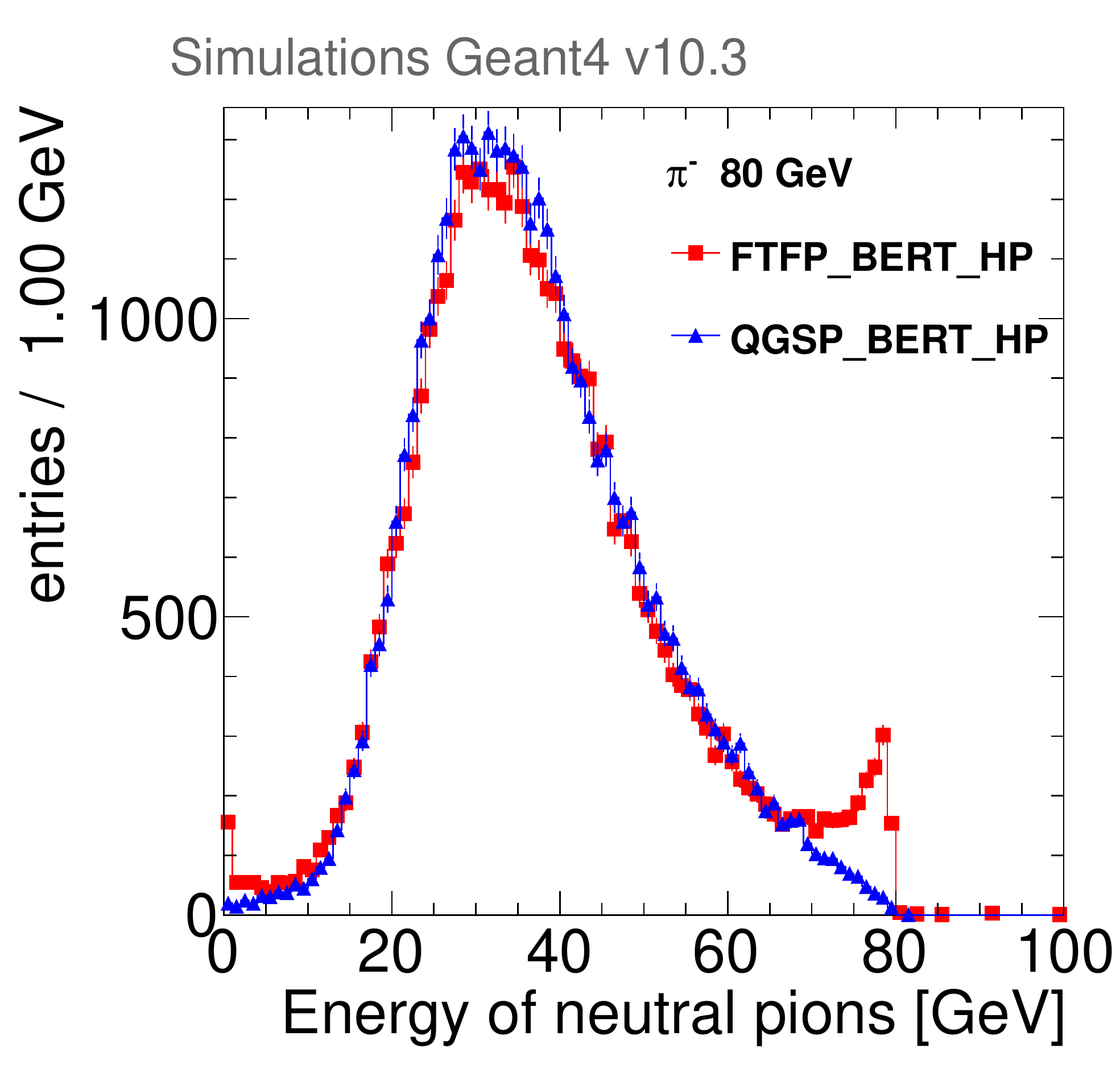}
\caption{\label{fig:epi0} Total energy of neutral pions produced in a hadronic shower initiated by $\pi^{-}$ with energy of 10~GeV (left), 40~GeV (middle) and 80~GeV (right) as simulated using FTFP\_BERT\_HP (red) or QGSP\_BERT\_HP (blue) physics lists of Geant4 version 10.3. }
\end{figure}

The generator-level information about propagated particles is managed by the dedicated classes and collections in the Geant4 package. The information kept on an event-by-event basis includes particle type, initial four-momentum and also links to particle parents and daughters. Two types of secondary particles were studied: neutral pions and neutrons. The neutral pions born in the first and subsequent inelastic interactions decay immediately into gammas that in turn produce electromagnetic subshowers within a hadronic shower. The energy spectra of generated neutral pions is shown in figure \ref{fig:Epi0} for two physics lists and three initial pion energies. The difference between FTF and QGS models can be seen for 40 and 80~GeV.  The total energy of generated $\pi^{0}$s contains information about electromagnetic fraction in an event. The distributions of the sum of $\pi^{0}$s' energies are shown in figure \ref{fig:epi0}. Hereinafter this sum is denoted as  $E_{\pi^{0}}$ or "energy of neutral pions".

\begin{figure}[htbp]
\centering 
\includegraphics[width=.45\textwidth]{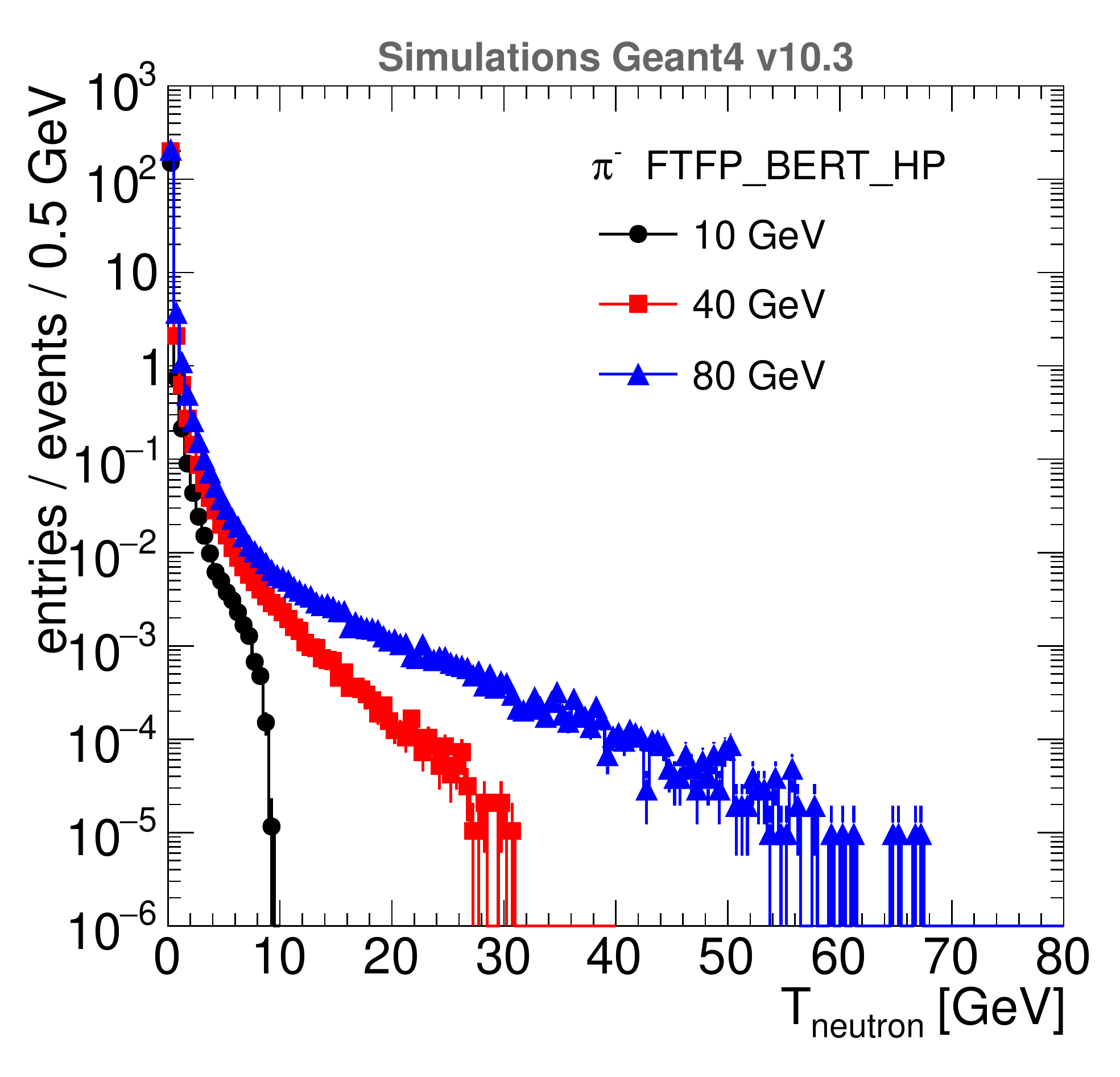}
\qquad
\includegraphics[width=.45\textwidth]{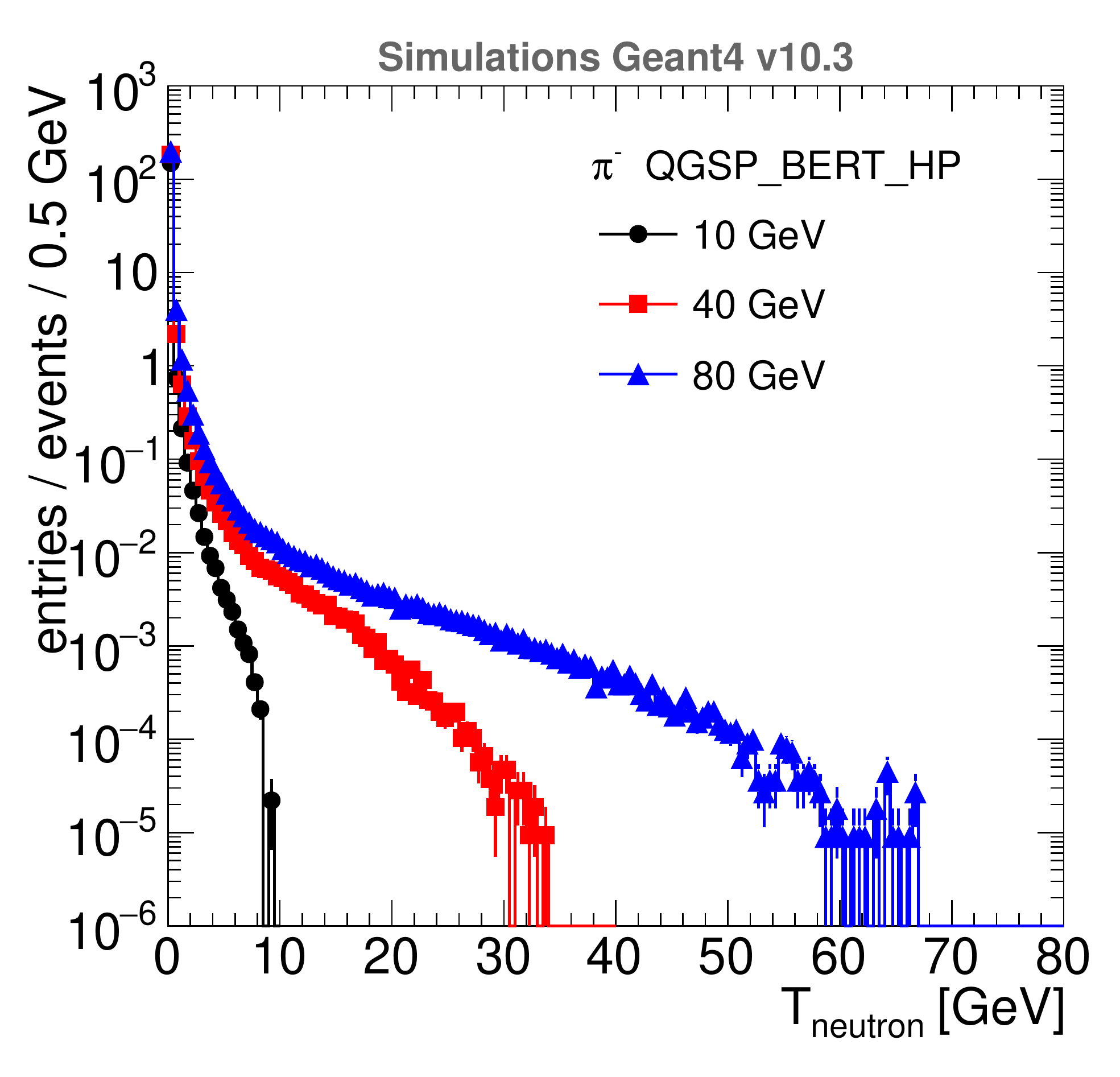}
\caption{\label{fig:Tneu} Kinetic energies of neutrons in a hadronic shower initiated by $\pi^{-}$ with energy 10~GeV (black), 40~GeV (red) and 80~GeV (blue) as simulated using FTFP\_BERT\_HP (left) or QGSP\_BERT\_HP (right) physics lists of Geant4 version 10.3.}
\end{figure}

\begin{figure}[htbp]
\centering 
\includegraphics[width=.32\textwidth]{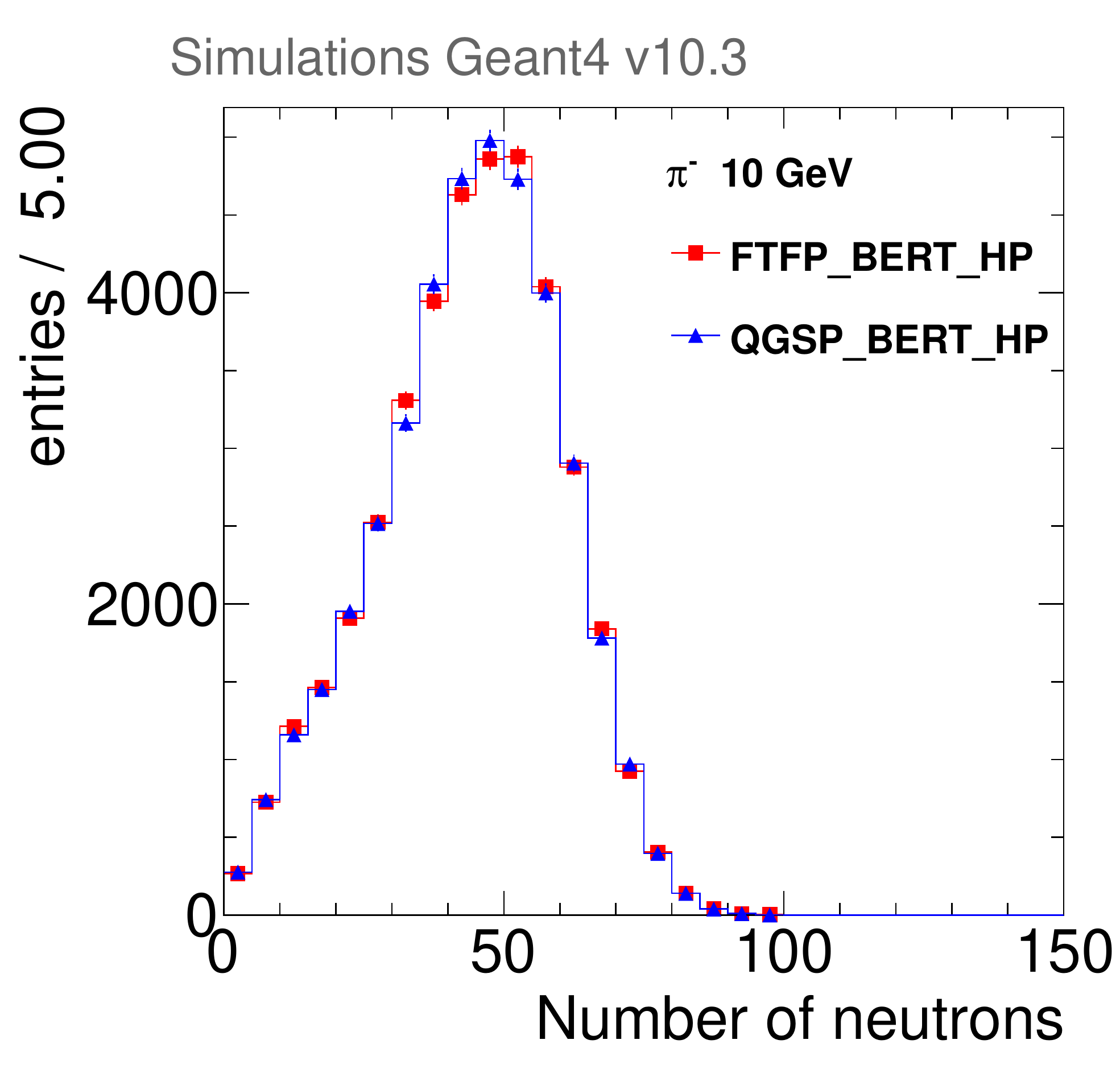}
\includegraphics[width=.32\textwidth]{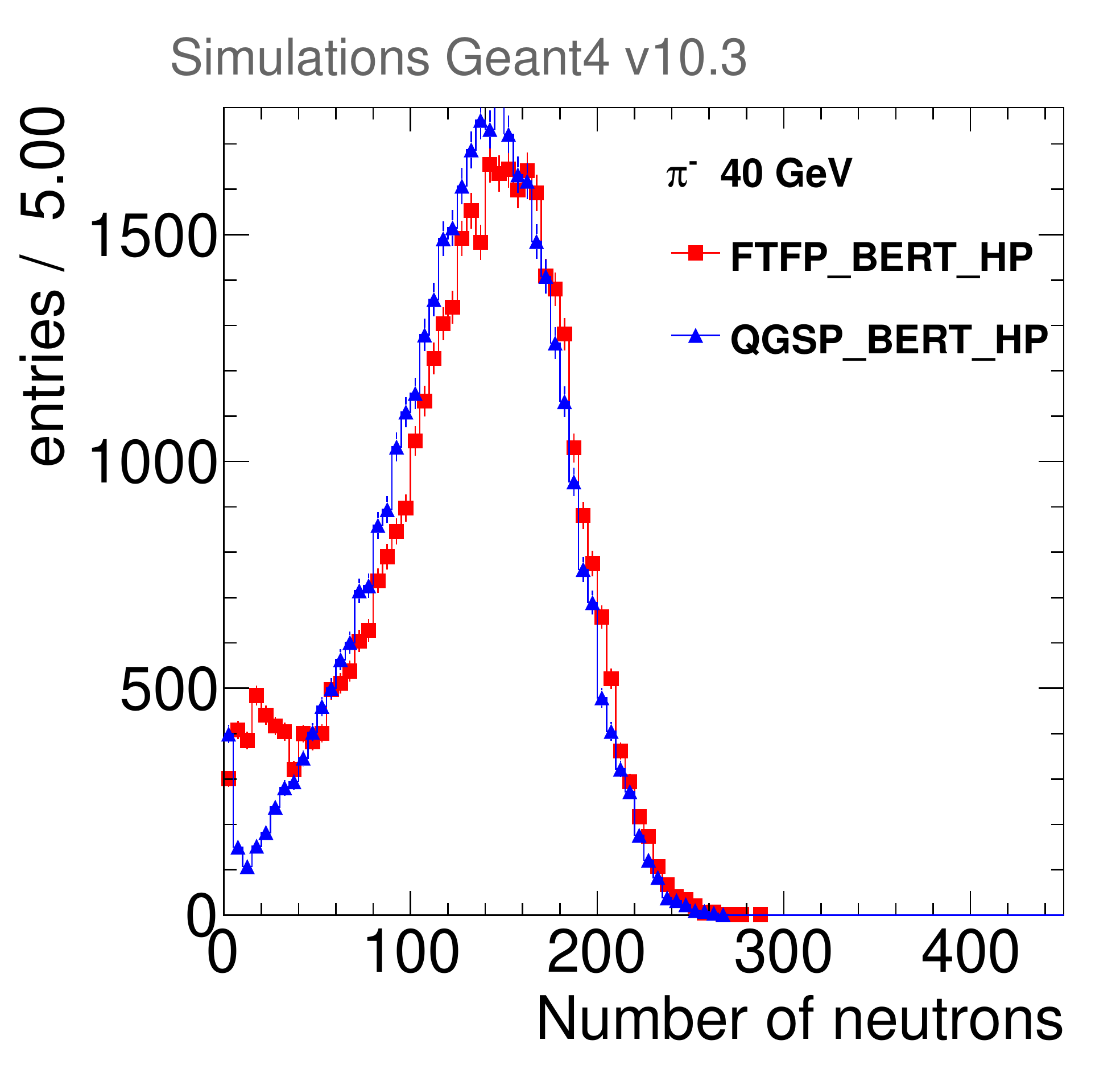}
\includegraphics[width=.32\textwidth]{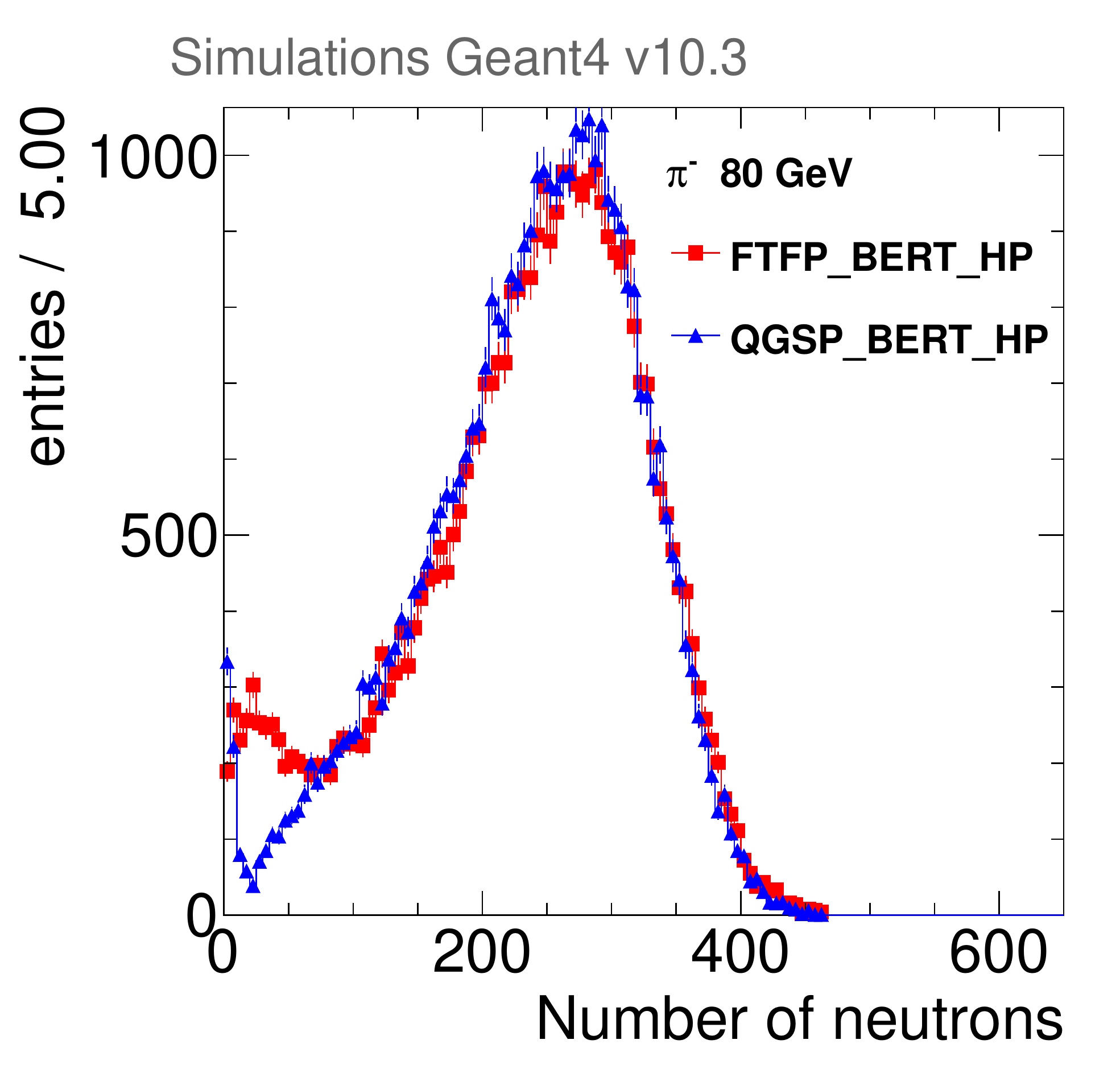}
\caption{\label{fig:nneu} Number of neutrons in a hadronic shower initiated by $\pi^{-}$ with energy of 10~GeV (left), 40~GeV (middle) and 80~GeV (right) as simulated using FTFP\_BERT\_HP (red) or QGSP\_BERT\_HP (blue) physics lists of Geant4 version 10.3.}
\end{figure}

The extraction of properties of secondary neutrons is not so straightforward as for neutral pions. While neutral pions decay immediately, the neutrons travel through calorimeter exhibiting different interactions including elastic scatterings. To avoid double counting of the same neutrons, those of them are excluded, which have only one parent, which is also neutron. The remaining neutrons are counted and this number is kept on an event-by-event basis for the further analysis of secondaries. The spectra of kinetic energies of the counted secondary neutrons and the distributions of their number are shown in figures \ref{fig:Tneu} and \ref{fig:nneu}, respectively. The similarity of spectra as well as distributions at 10~GeV is due to domination of Bertini model in this energy range for both physics lists.

\section{Calorimetric observables}
\label{sec:calo_obs}

The standard calorimetric observables, such as total number of hits and reconstructed energy, are calculated using all hits produced in the calorimeter including the primary track hits. Another observables defined for this study are calculated from the shower hits only. The longitudinal position of the first inelastic interaction (shower start layer) is identified on an event-by-event basis. No additional clustering is applied to the simulated single particle events and the hits from and beyond the identified shower start layer are all considered to belong to the hadronic shower. For the event preselection applied as discussed in section \ref{sec:conditions}, the overwhelming majority of hits belong to the shower hits. The following observables are calculated for each event:

\begin{itemize}
\item \textbf{Total number of hits}, $N_{\mathrm{hits}}$.
\item \textbf{Reconstructed energy}, $E_{\mathrm{reco}} = \sum_{i=1}^{N_{\mathrm{hits}}}{e_i}$, where $e_i$ is the measured energy of the $i$-th hit in units of MIP. 
\item \textbf{Number of shower hits},  $N_{\mathrm{sh}}$, counted beyond the identified shower start.  
\item \textbf{Number of isolated hits in a shower}, $N_{\mathrm{iso}}$. The hit is called isolated if it has no neighbour hits in a cube of 3$\times$3$\times$3 cells around it.
\item \textbf{Number of track hits in a shower}, $N_{\mathrm{trk}}$. Track hit is defined as a hit within a shower, which has 2 in-line neighbours and amplitude less than 5~MIP.
\item \textbf{Mean shower hit energy}, $\left<e_{\mathrm{hit}}\right> = \frac{1}{N_{\mathrm{sh}}} \sum_{i=1}^{N_{\mathrm{sh}}}{e_i}$.
\item \textbf{Shower radius}, $R_{\mathrm{sh}} = \frac{1}{\sum_{i=1}^{N_{\mathrm{sh}}}{e_{i}}} \sum_{i=1}^{N_{\mathrm{sh}}}{e_{i} \cdot r_{i}}$, where $r_i = \sqrt{(x_i - X_{CoG})^2 + (y_i - Y_{CoG})^2}$ is the radial distance of the $i$-th hit with coordinates ($x_i$, $y_i$) from the shower radial centre of gravity with coordinates ($X_{CoG}$,$Y_{CoG}$). The coordinates of the radial center of gravity are defined as a weighted sums $X_{\mathrm{CoG}} = \frac{\sum_{i=1}^{N_{\mathrm{hits}}}{e_{i} \cdot x_{i}}}{\sum_{i=1}^{N_{\mathrm{hits}}}{e_{i}}}$ and $Y_{\mathrm{CoG}} = \frac{\sum_{i=1}^{N_{\mathrm{hits}}}{e_{i} \cdot y_{i}}}{\sum_{i=1}^{N_{\mathrm{hits}}}{e_{i}}}$.
\item \textbf{Longitudinal shower centre of gravity}, $Z_{\mathrm{CoG}} = \frac{\sum_{i=1}^{N_{\mathrm{sh}}}{e_{i} \cdot (z_{i} - z_{\mathrm{start}})}}{\sum_{i=1}^{N_{\mathrm{sh}}}{e_{i}}}$, where $z_i$  is the hit longitudinal coordinate and $z_{\mathrm{start}}$ is the longitudinal coordinate of the identified position of the first inelastic interaction in units of $\lambda^{\mathrm{eff}}_{\mathrm{I}}$ (for the AHCAL, $\lambda^{\mathrm{eff}}_{\mathrm{I}} =$~226.5 mm, 0.118$\cdot \lambda^{\mathrm{eff}}_{\mathrm{I}}$/layer). 
\end{itemize}

\begin{figure}[htbp]
\centering 
\includegraphics[width=.32\textwidth]{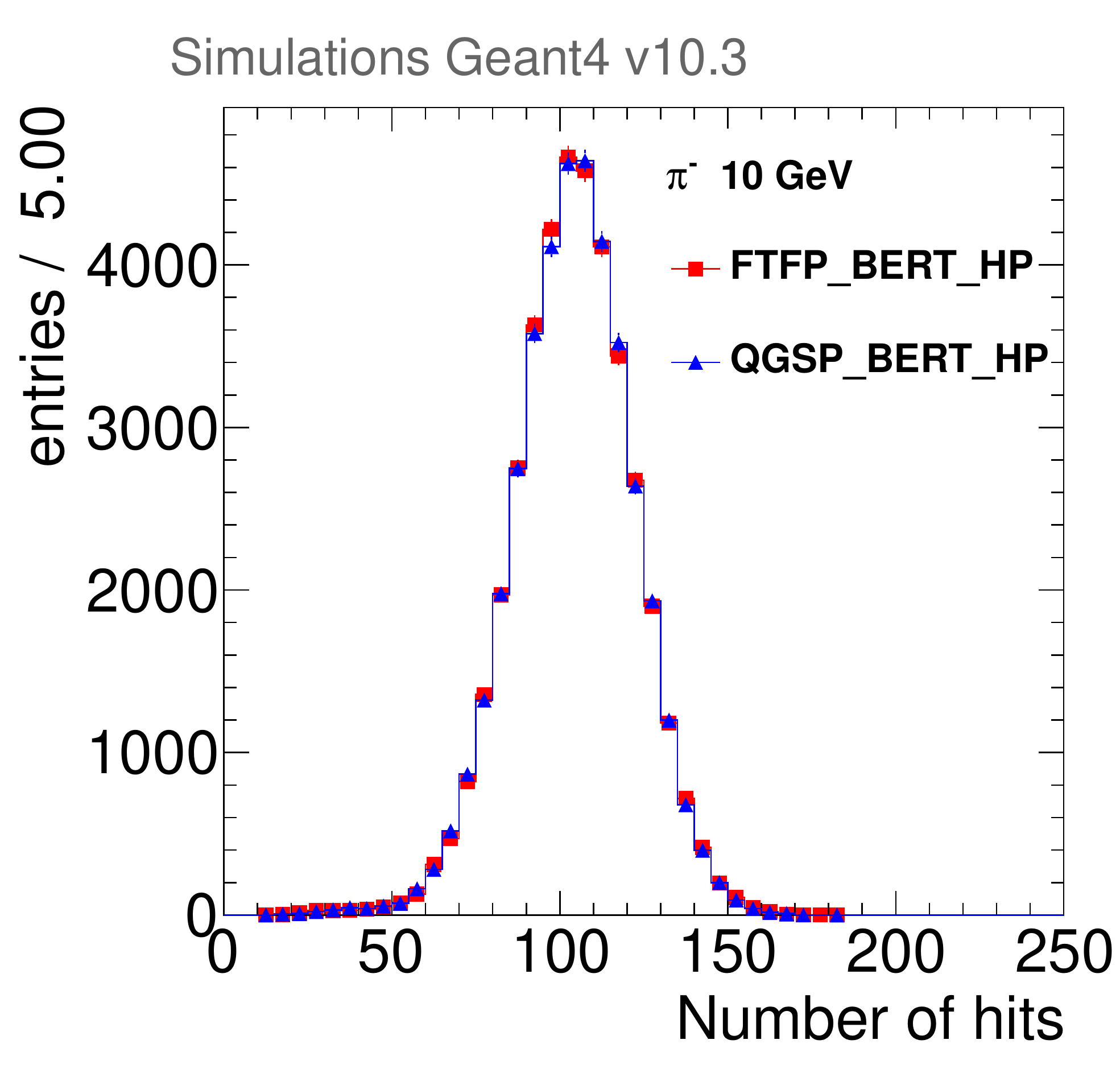}
\includegraphics[width=.32\textwidth]{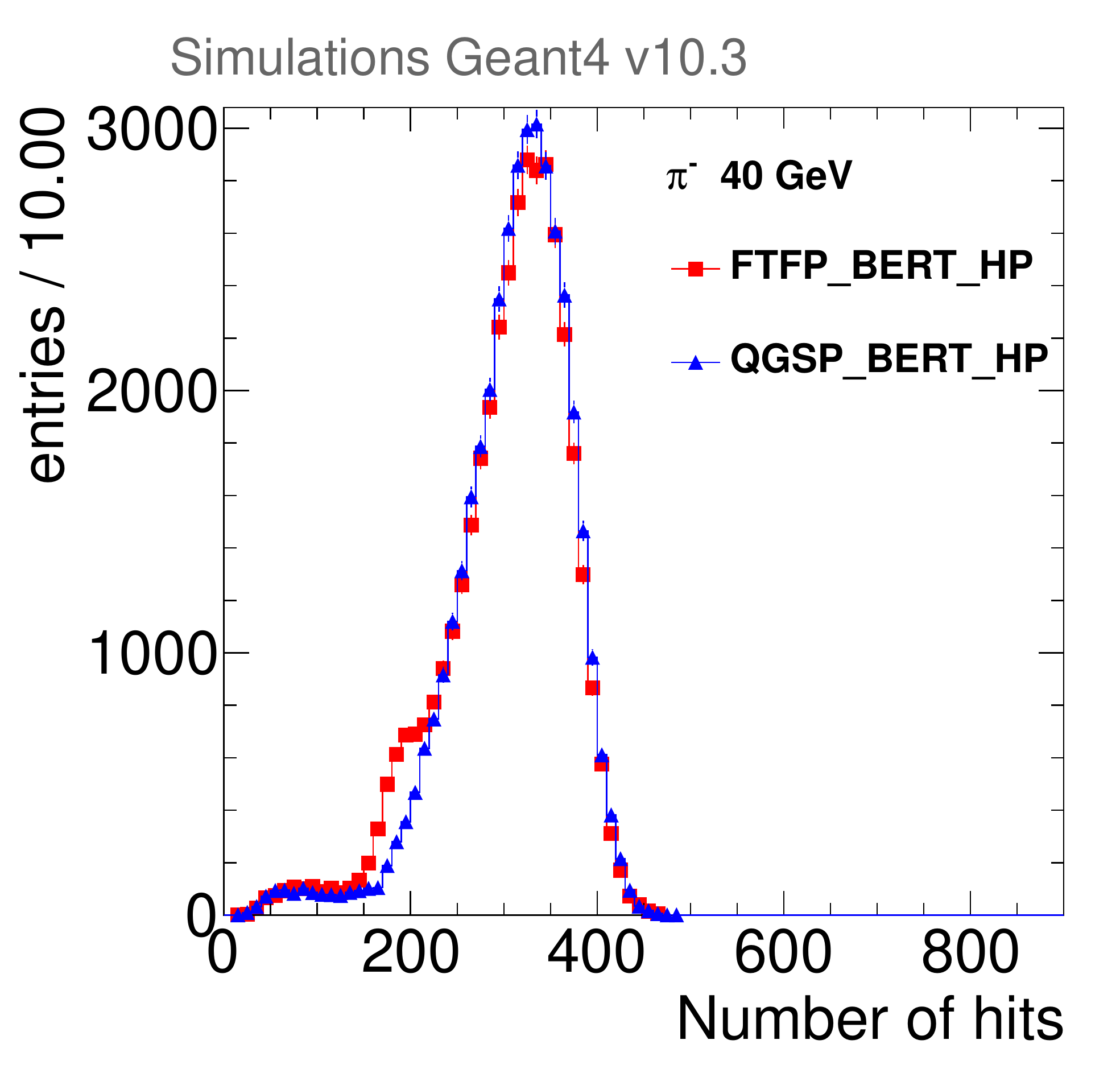}
\includegraphics[width=.32\textwidth]{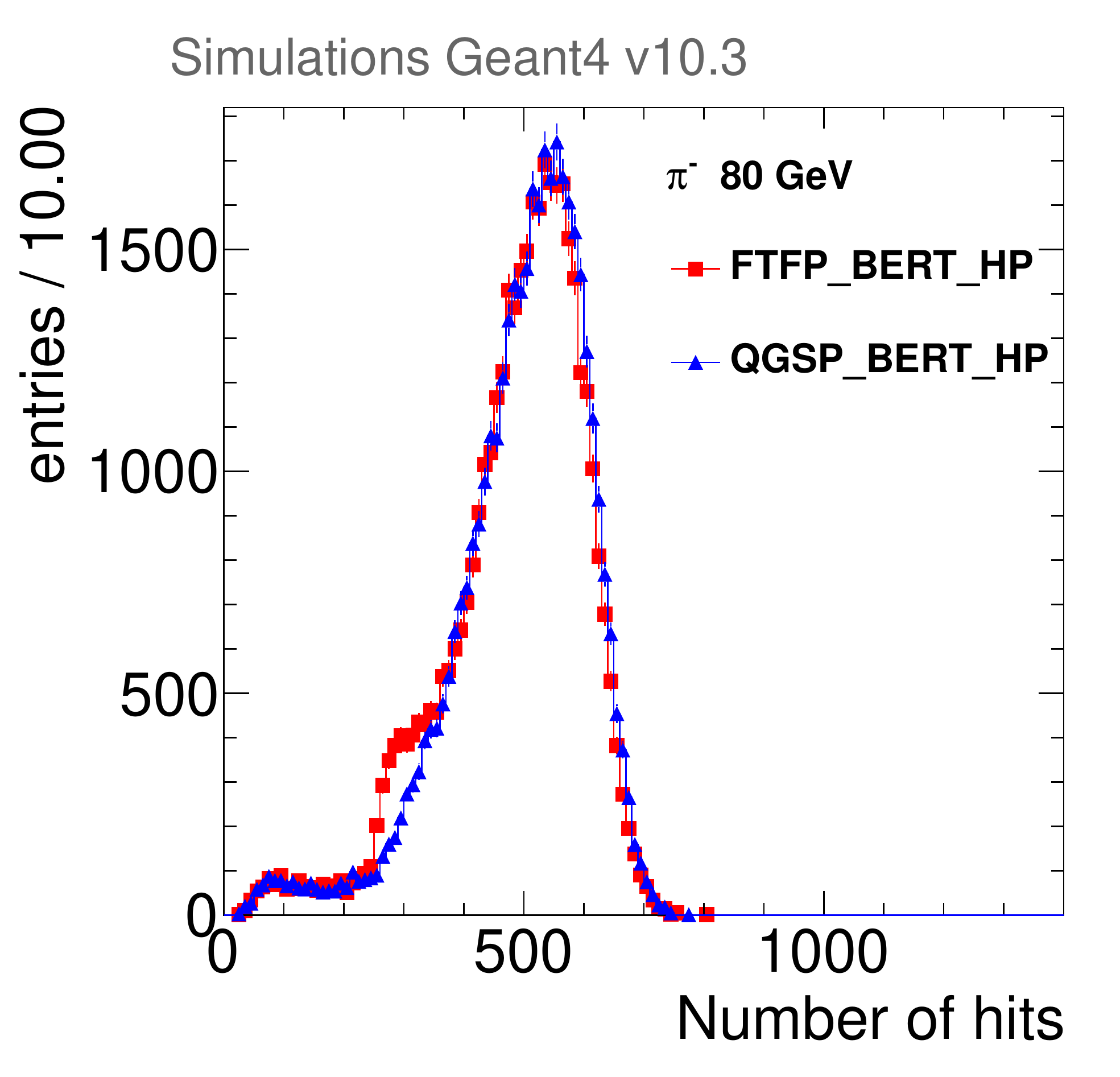}\\
\includegraphics[width=.32\textwidth]{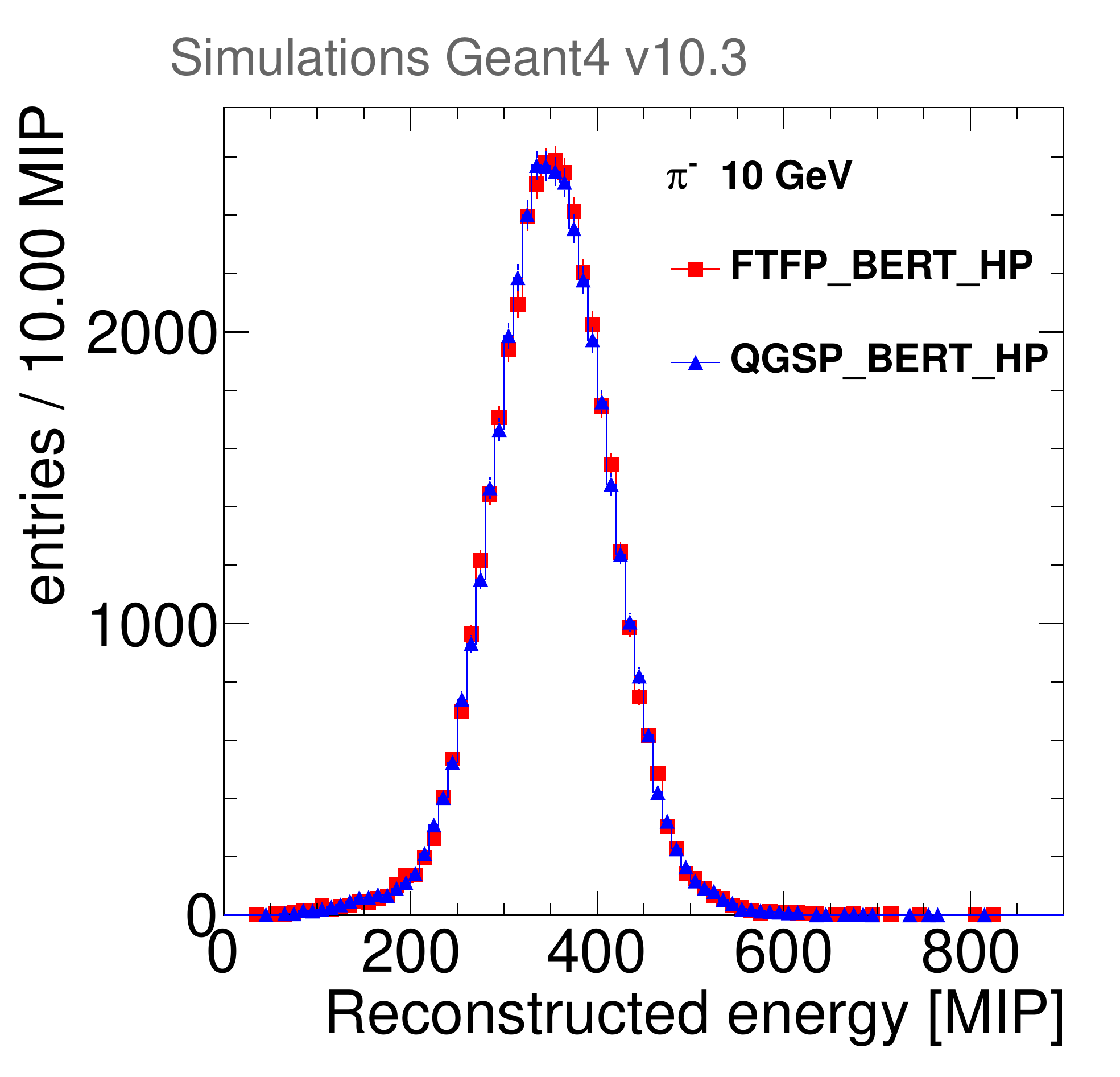}
\includegraphics[width=.32\textwidth]{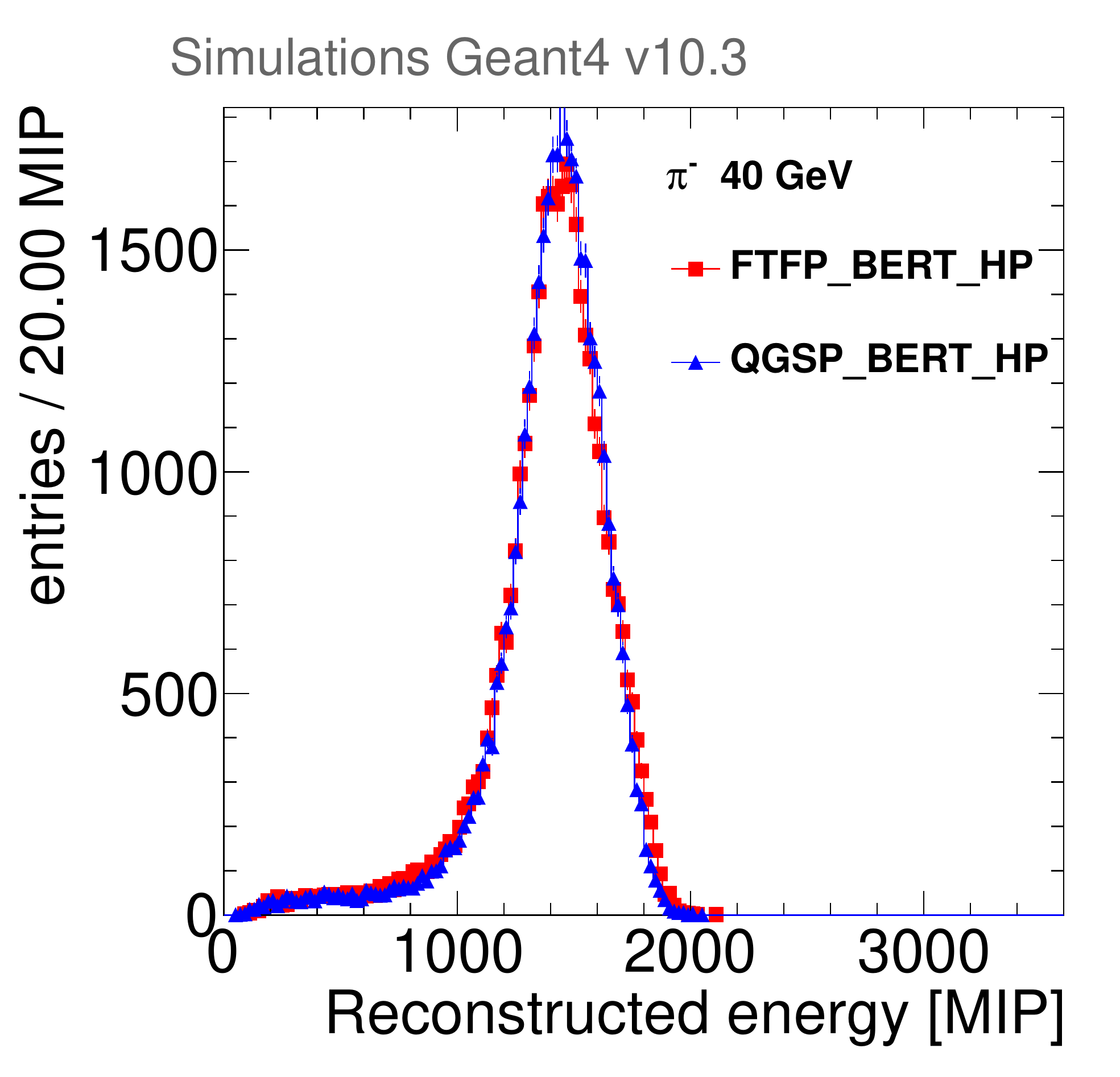}
\includegraphics[width=.32\textwidth]{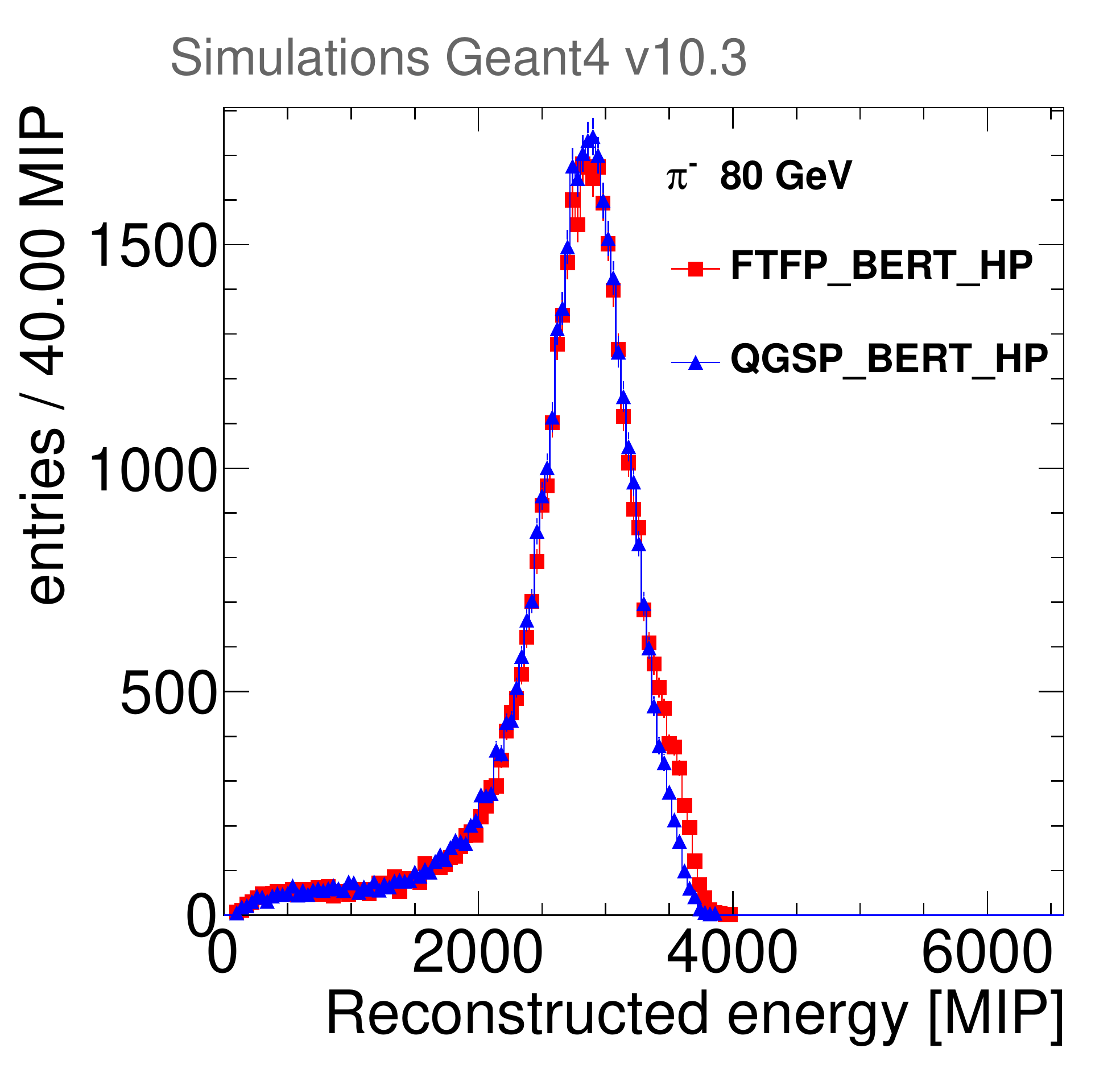}\\
\caption{\label{fig:obs_nhits_esum} Number of hits (upper row) and reconstructed energy (bottom row) of hadronic showers initiated by $\pi^{-}$ with energy of 10~GeV (left), 40~GeV (middle) and 80~GeV (right) as simulated using FTFP\_BERT\_HP (red) or QGSP\_BERT\_HP (blue) physics lists of Geant4 version 10.3.}
\end{figure}

\begin{figure}[htbp]
\centering 
\includegraphics[width=.32\textwidth]{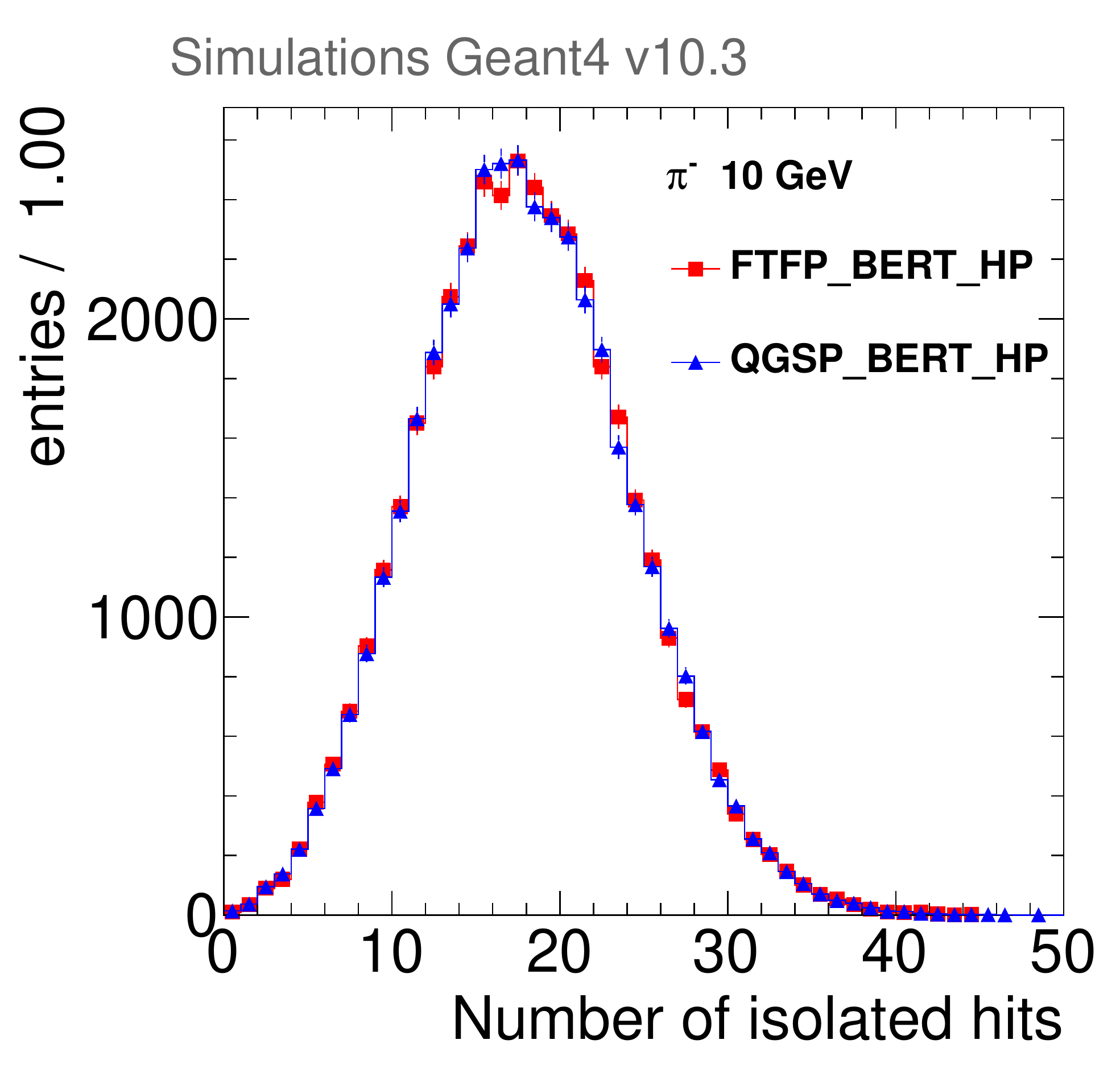}
\includegraphics[width=.32\textwidth]{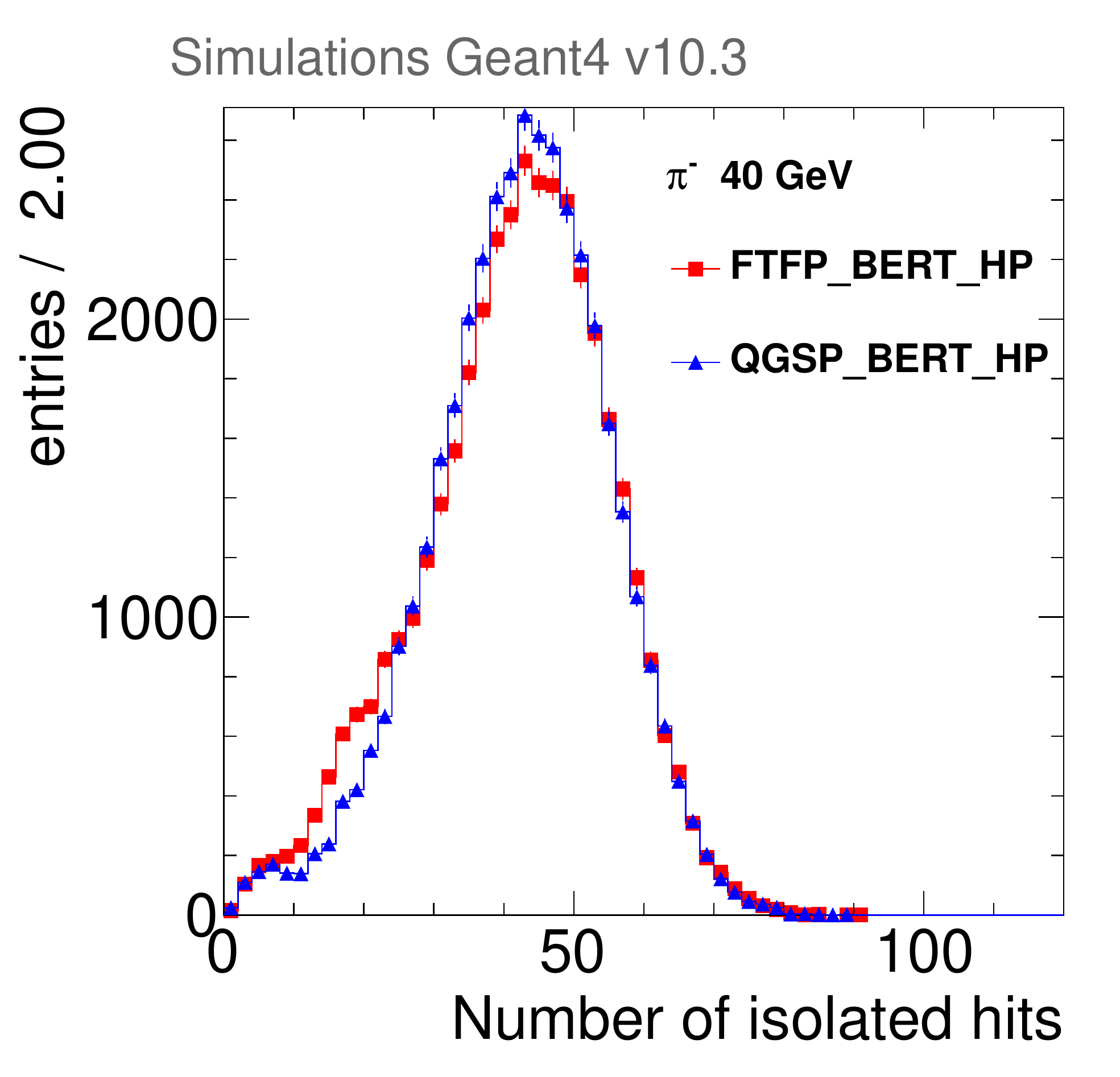}
\includegraphics[width=.32\textwidth]{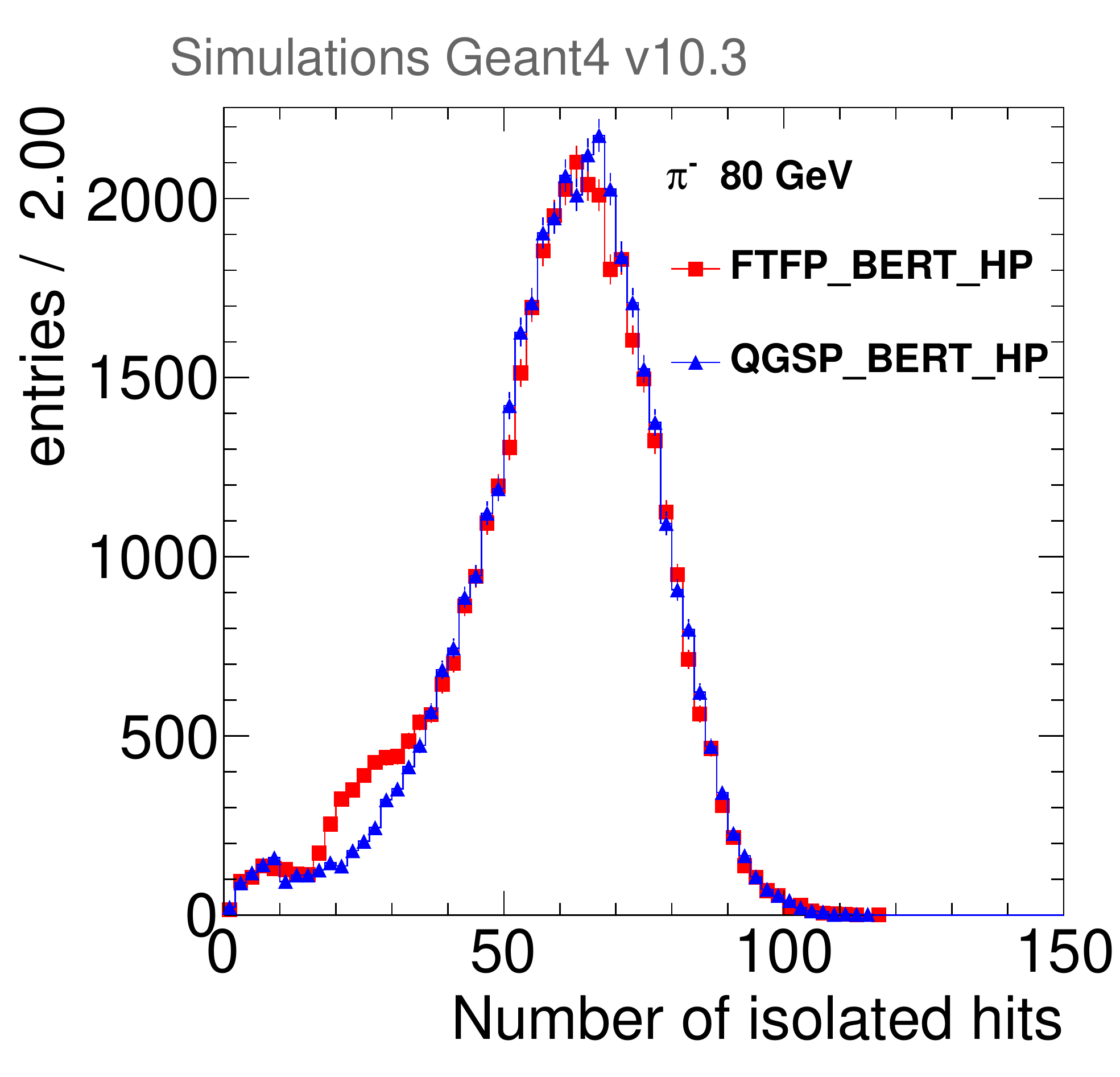}\\
\includegraphics[width=.32\textwidth]{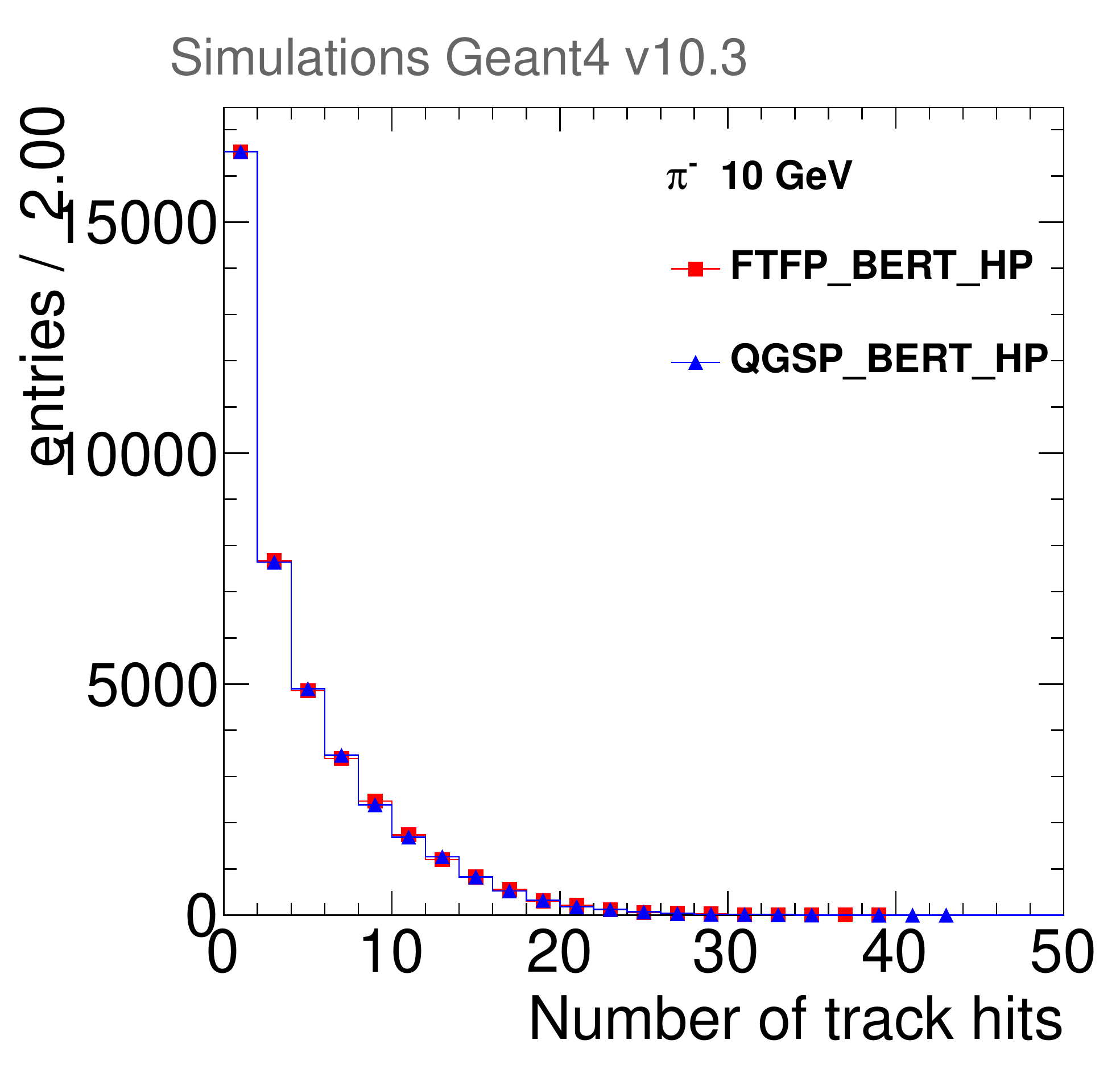}
\includegraphics[width=.32\textwidth]{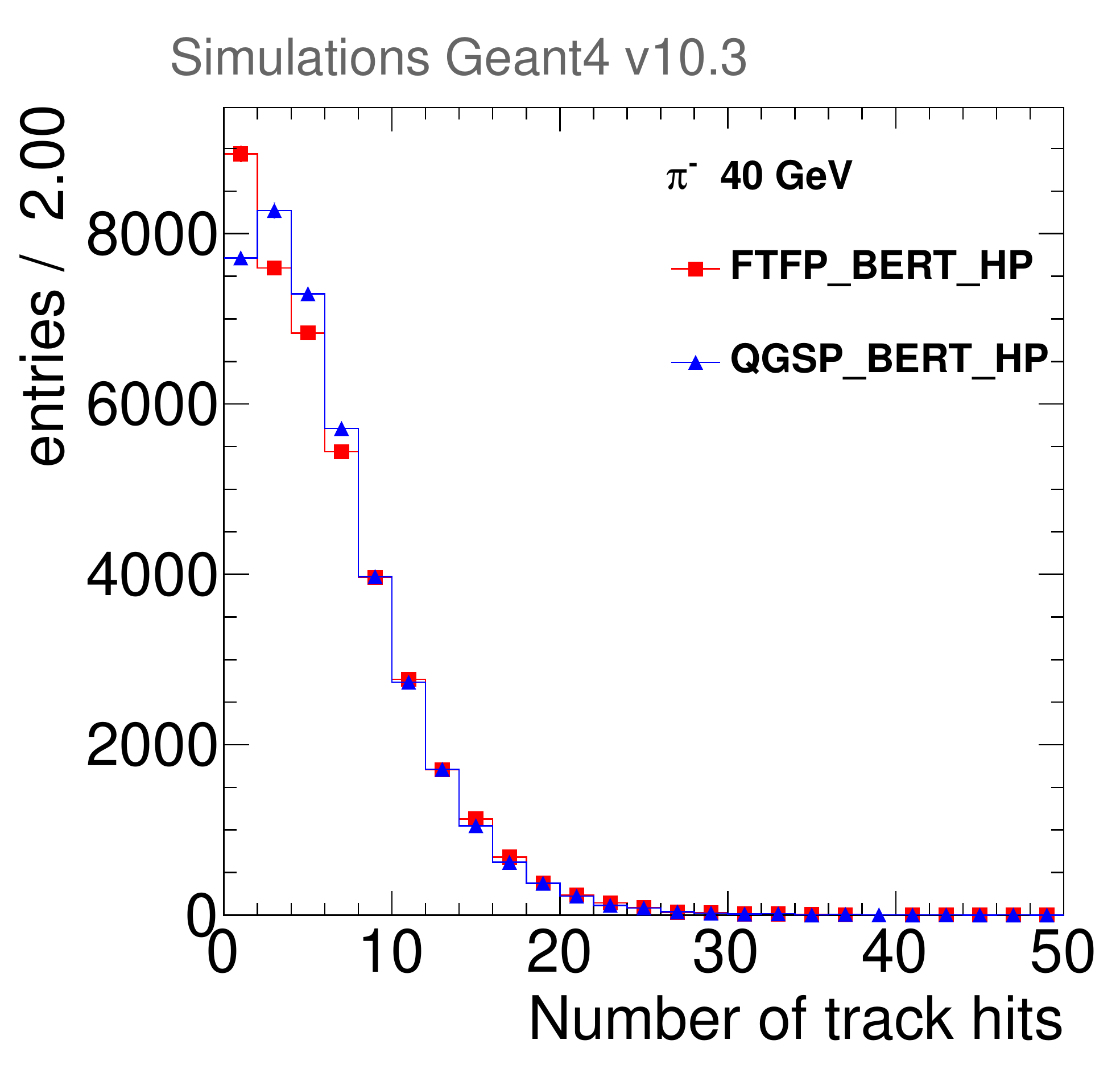}
\includegraphics[width=.32\textwidth]{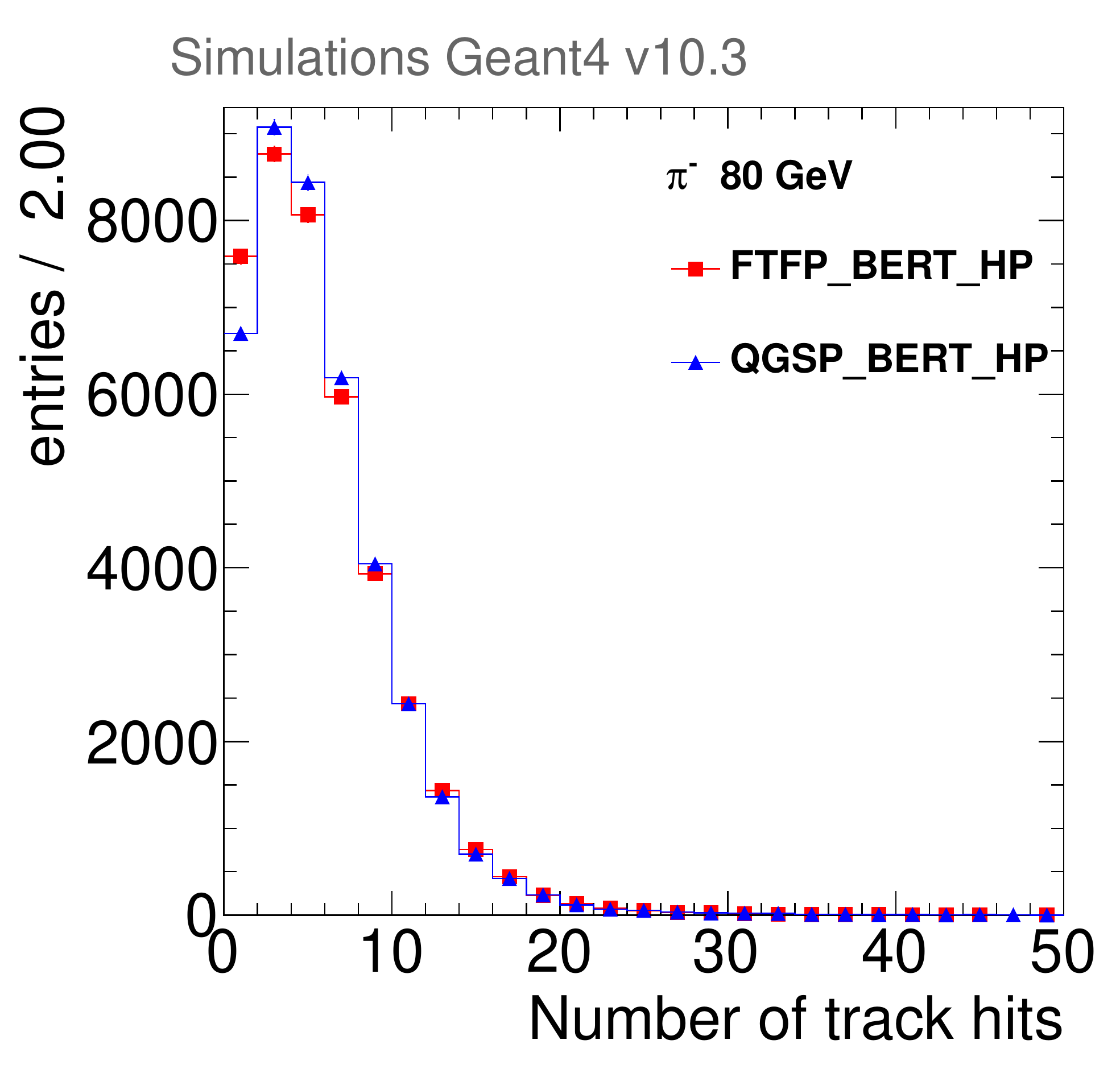}\\
\caption{\label{fig:obs_niso_ntrk} Number of isolated hits (upper row) and number of track hits within a hadronic shower initiated by $\pi^{-}$ with energy of 10~GeV (left), 40~GeV (middle) and 80~GeV (right) as simulated using FTFP\_BERT\_HP (red) or QGSP\_BERT\_HP (blue) physics lists of Geant4 version 10.3.}
\end{figure}

\begin{figure}[htbp]
\centering 
\includegraphics[width=.32\textwidth]{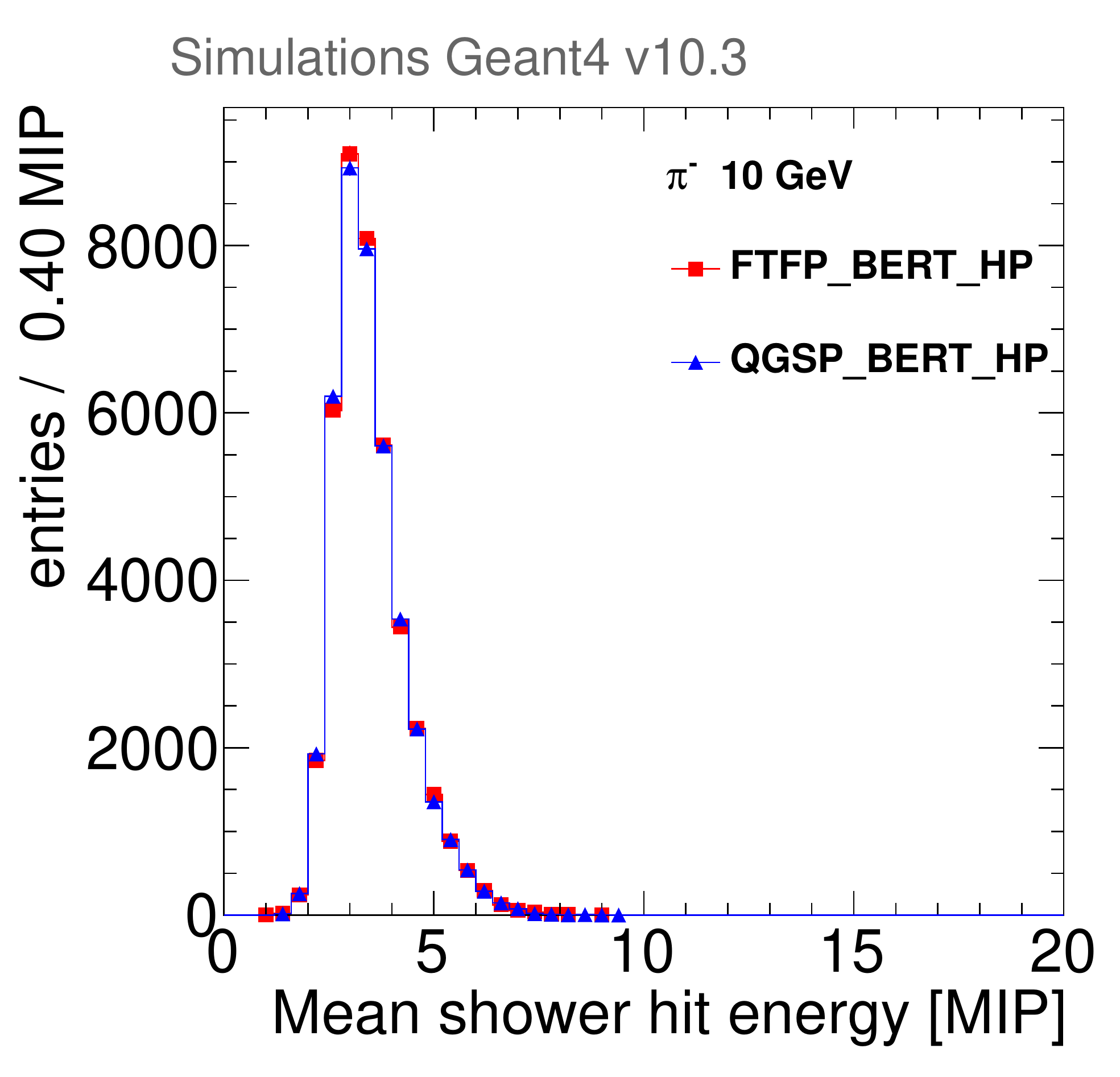}
\includegraphics[width=.32\textwidth]{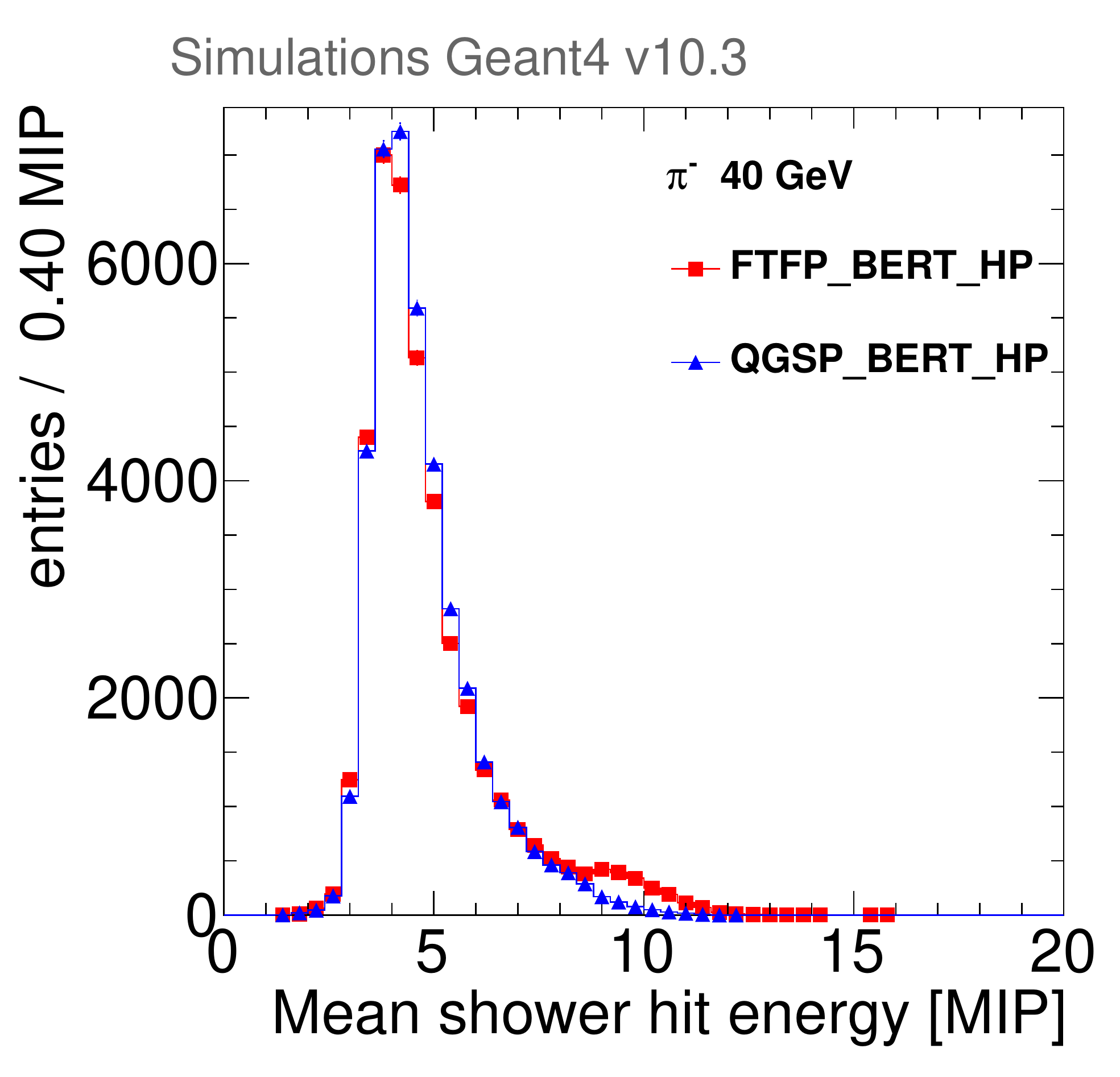}
\includegraphics[width=.32\textwidth]{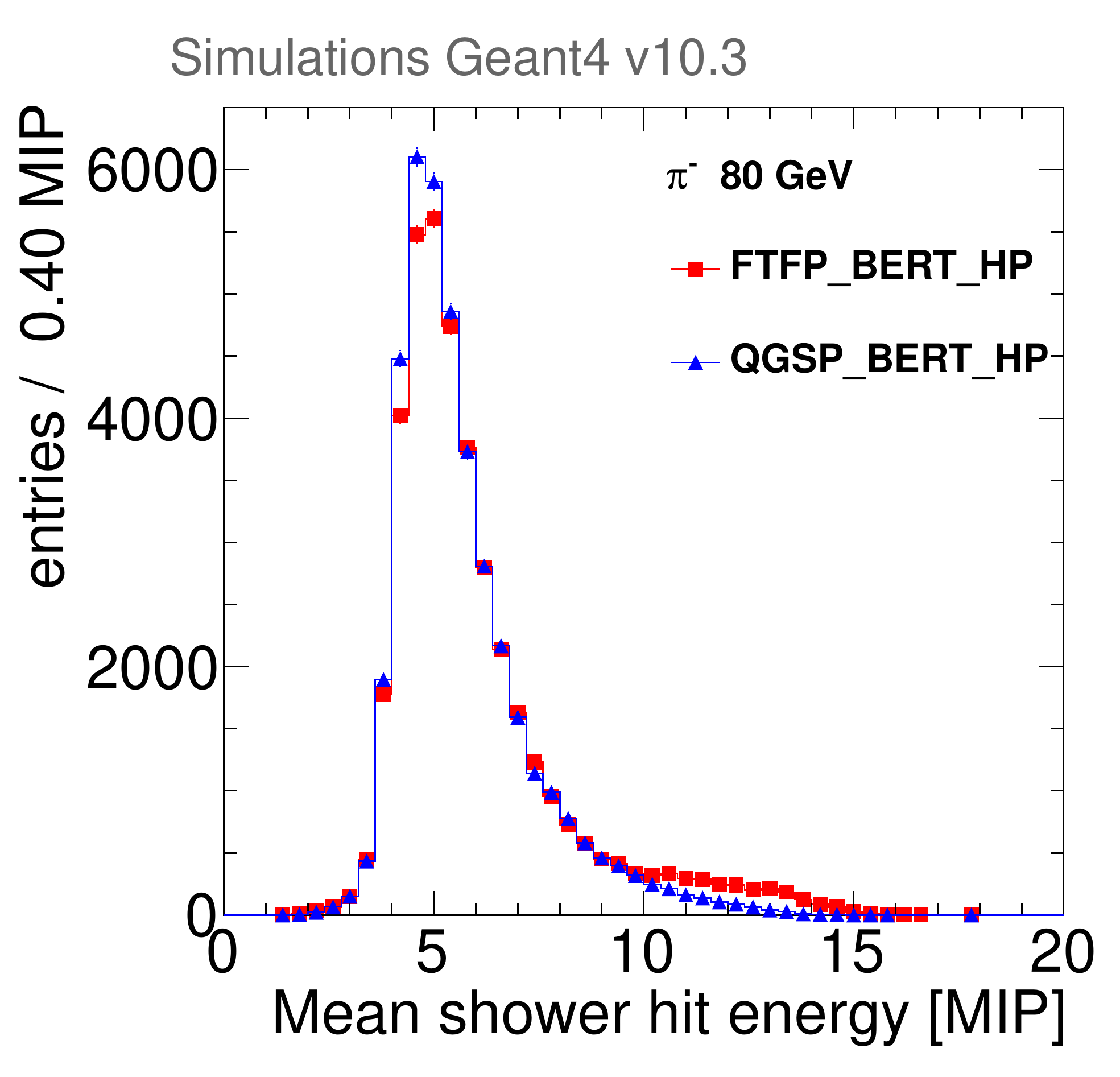}\\
\includegraphics[width=.32\textwidth]{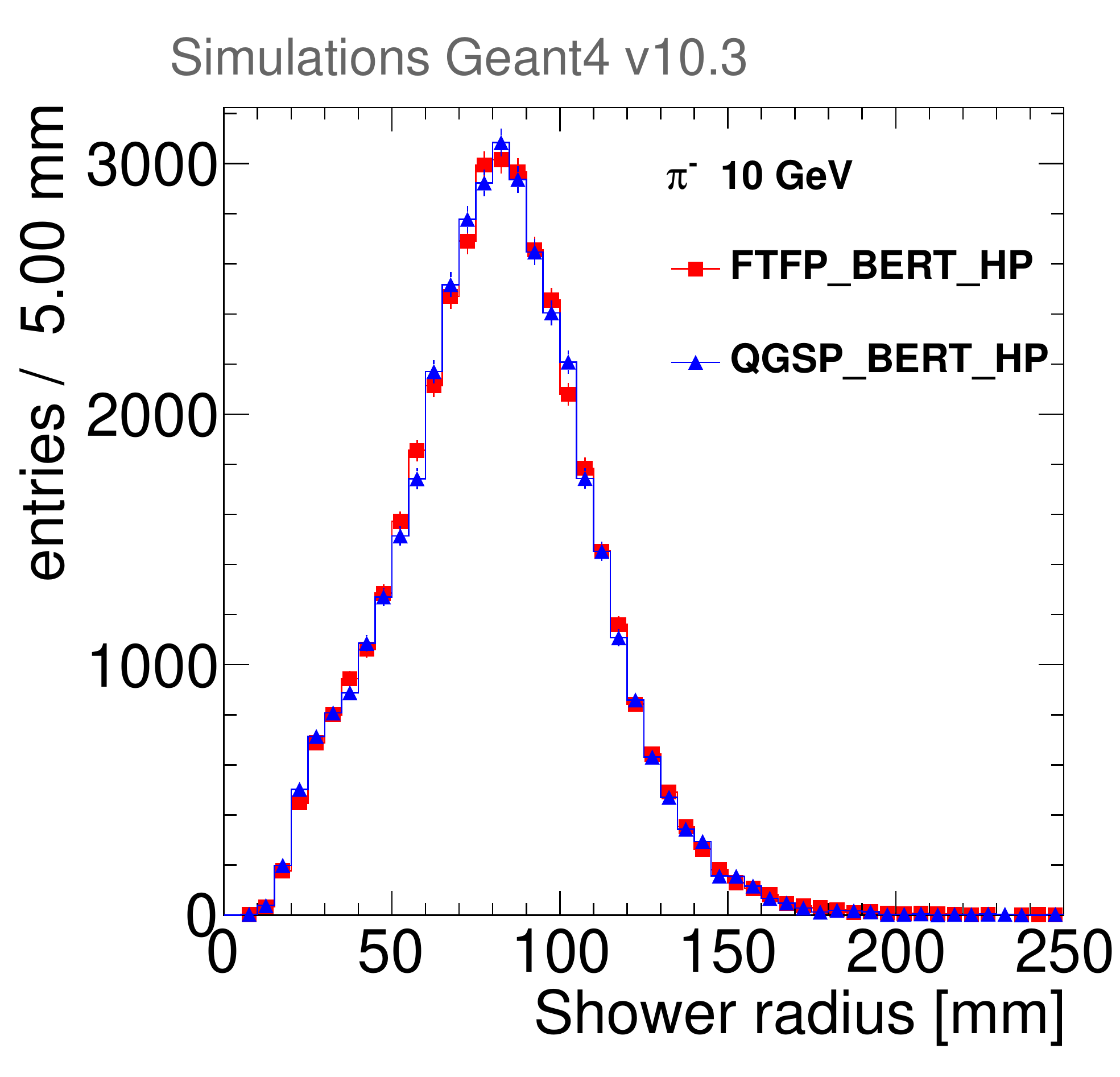}
\includegraphics[width=.32\textwidth]{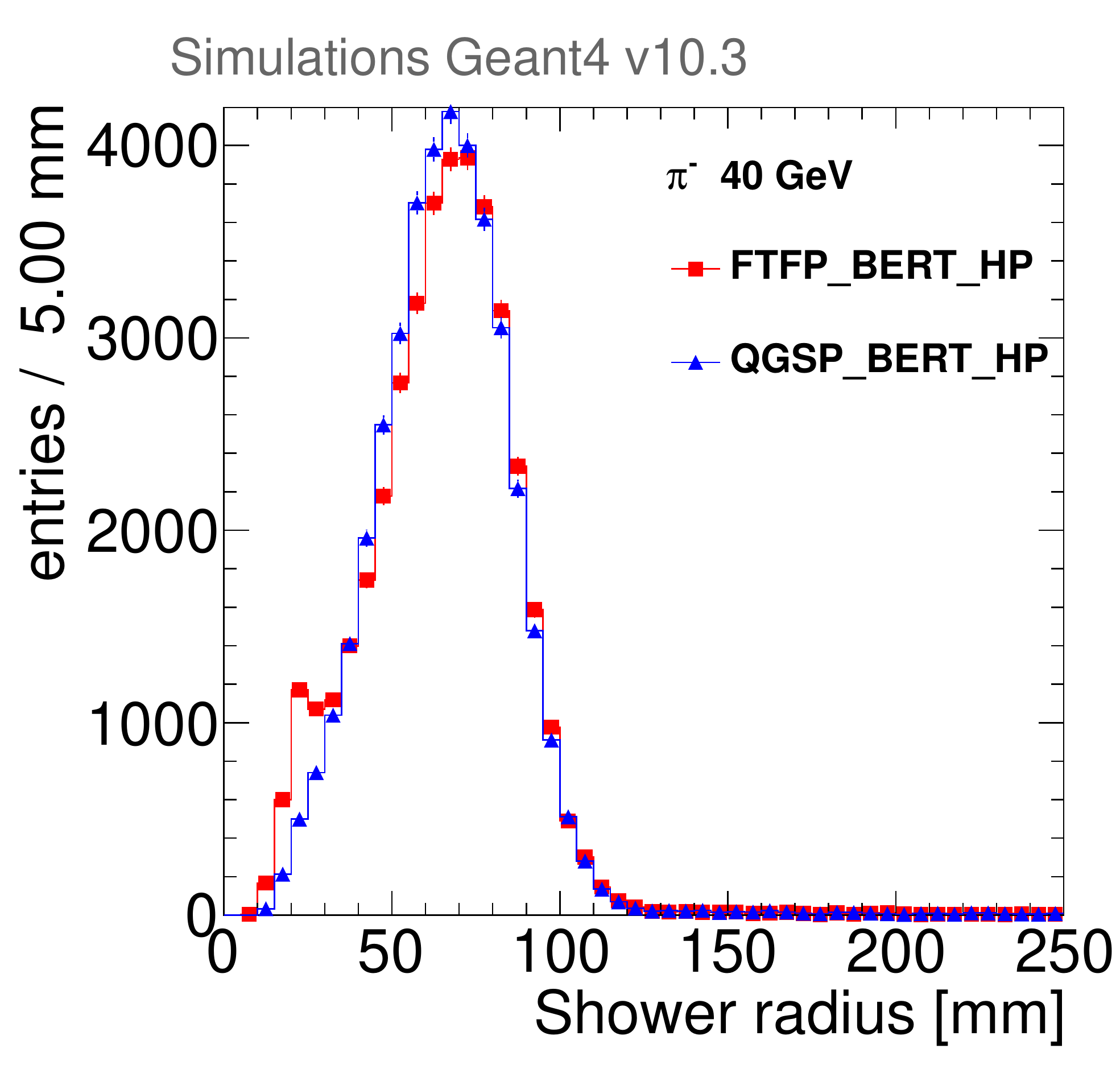}
\includegraphics[width=.32\textwidth]{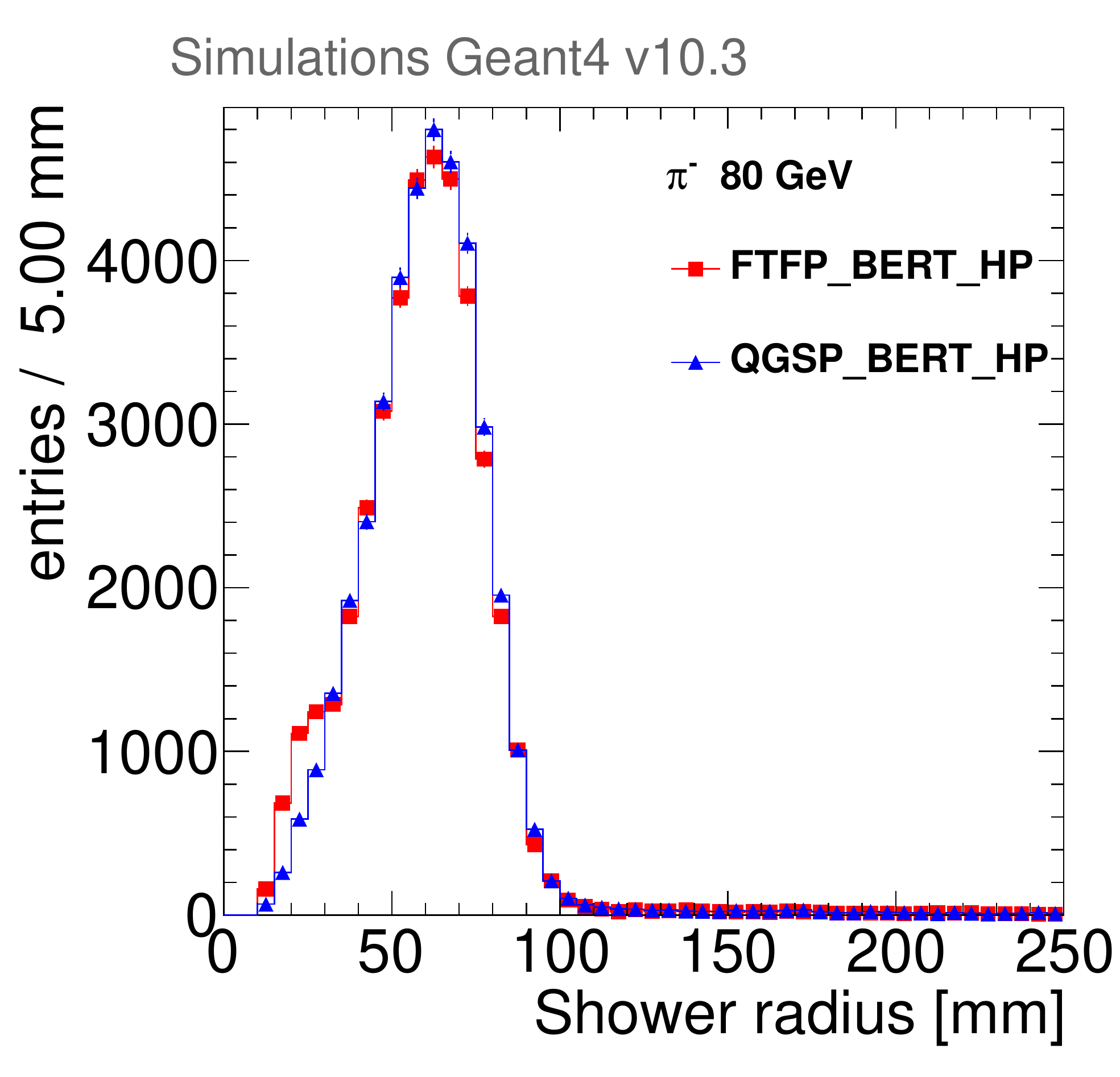}\\
\includegraphics[width=.32\textwidth]{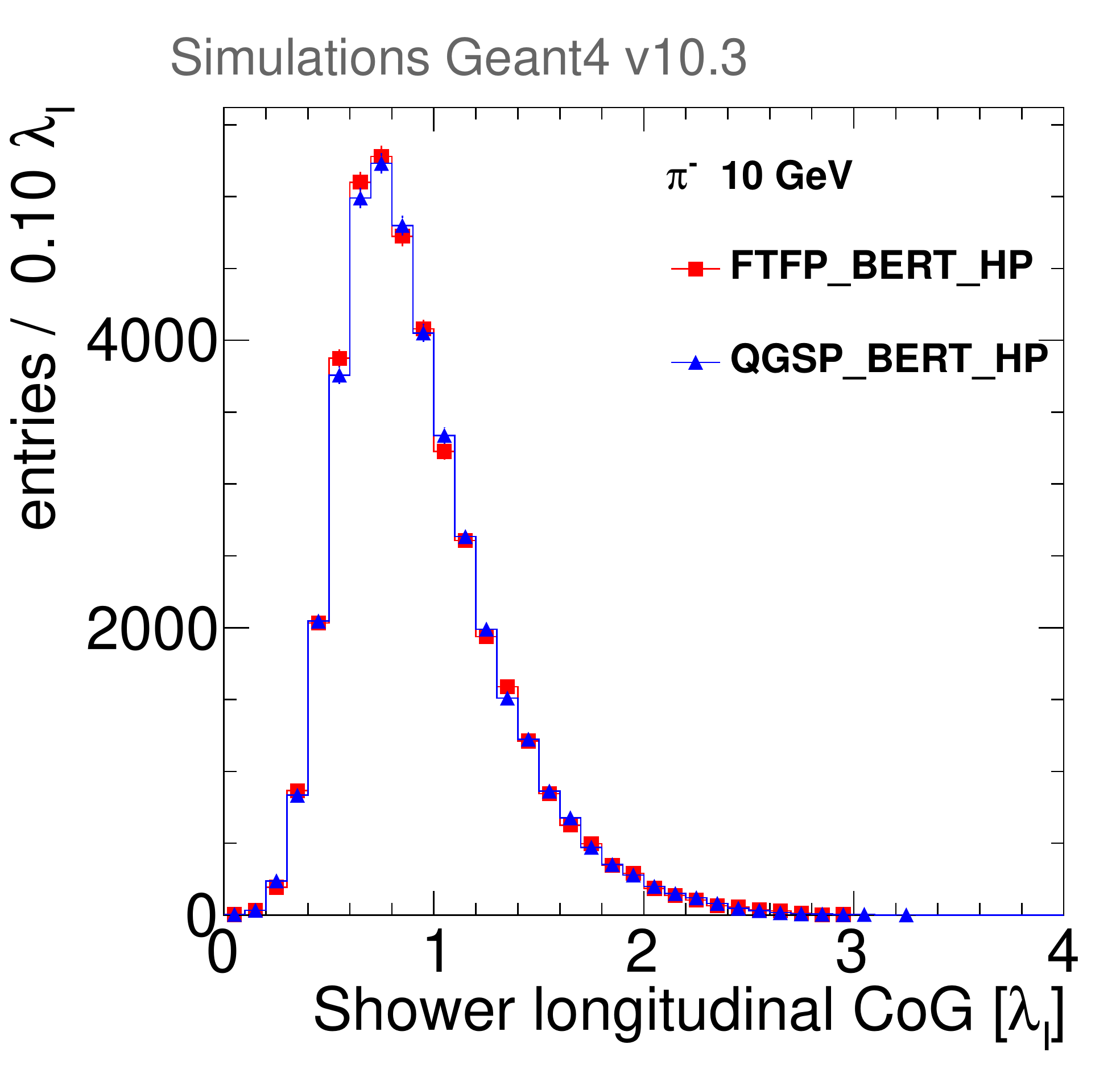}
\includegraphics[width=.32\textwidth]{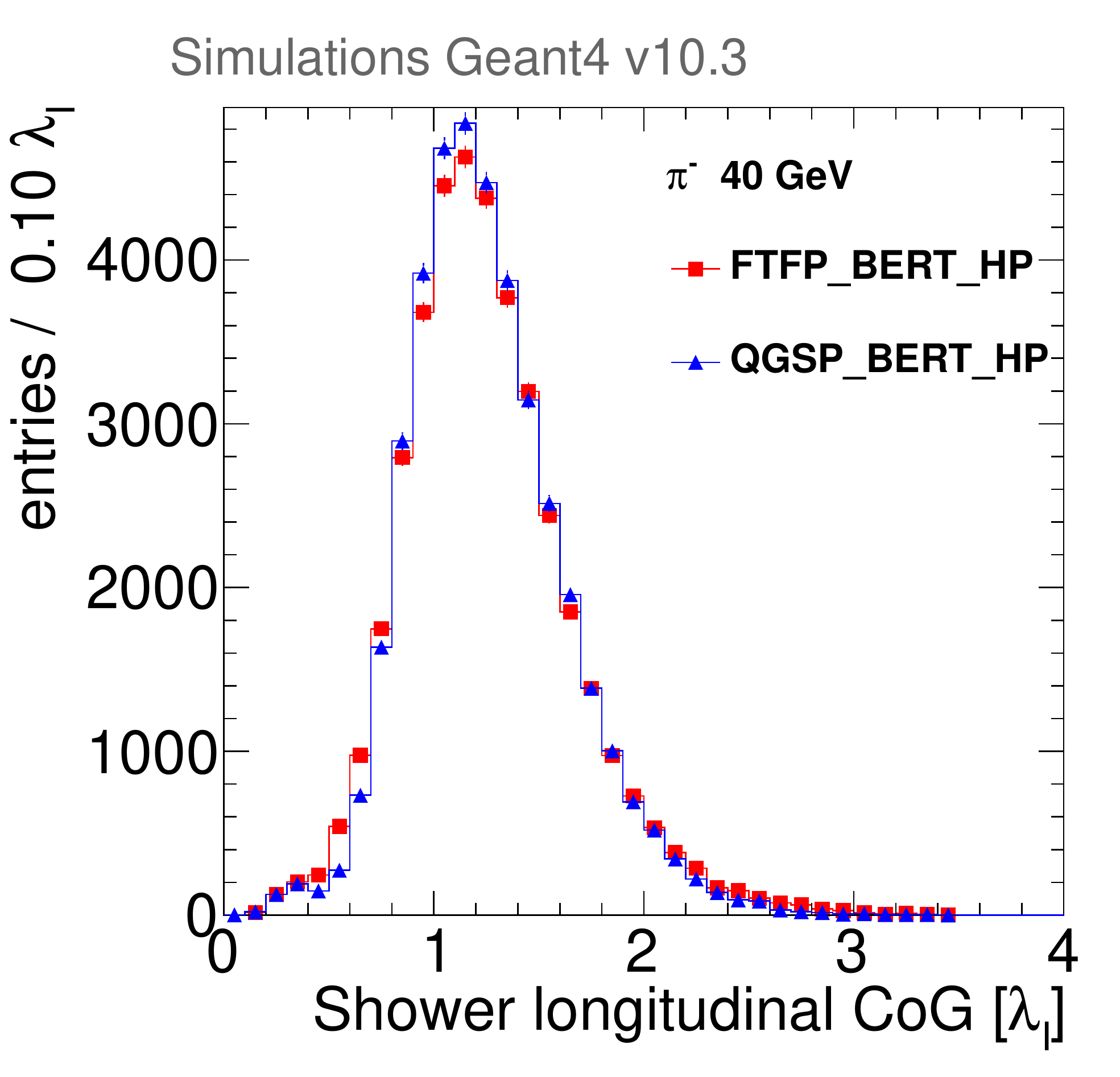}
\includegraphics[width=.32\textwidth]{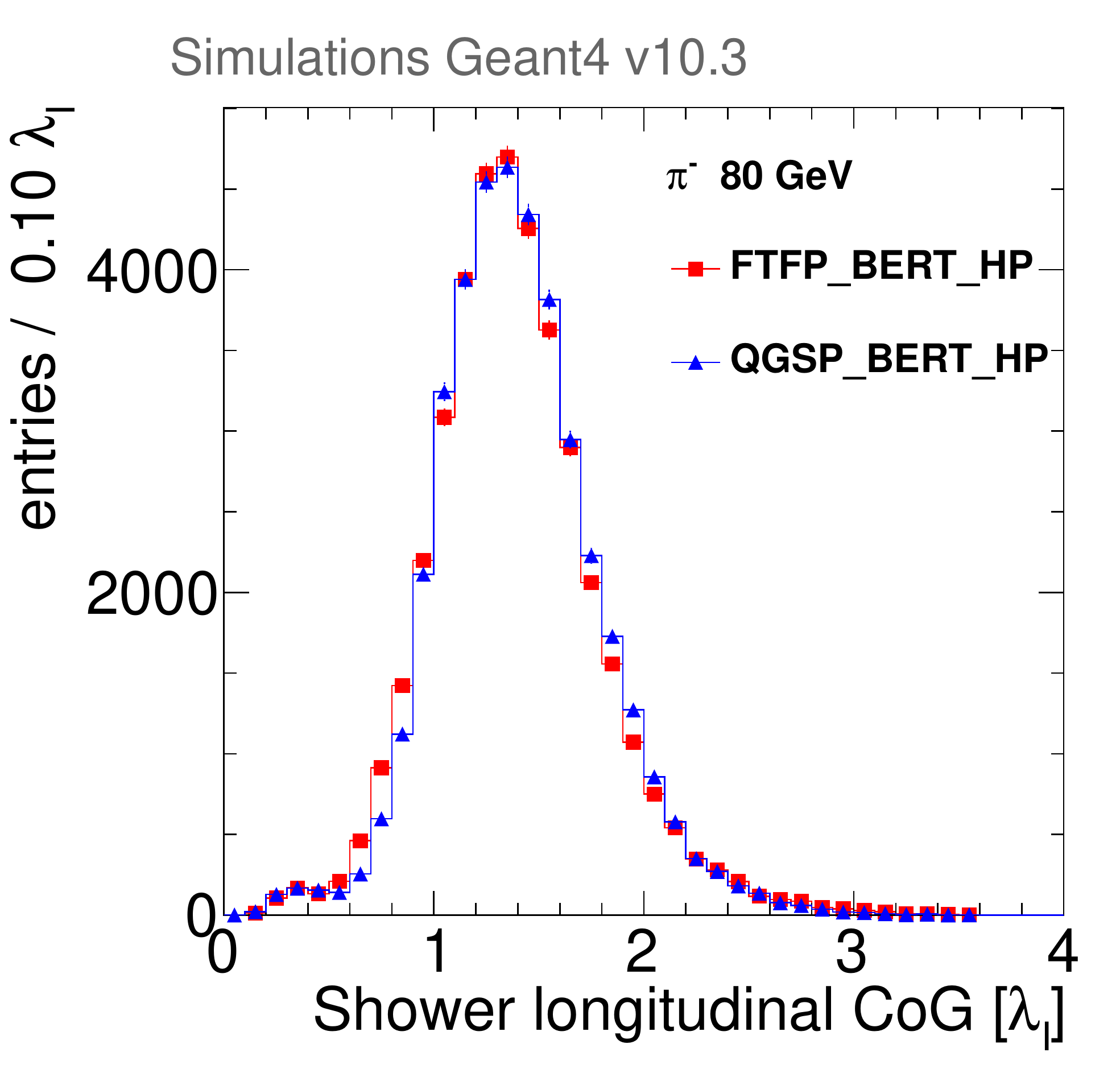}\\
\caption{\label{fig:obs_ehit_rad_zcog} Mean hit energy (upper row), shower radius (middle row) and longitudinal centre of gravity (bottom row) for hadronic showers initiated by $\pi^{-}$ with energy of 10~GeV (left), 40~GeV (middle) and 80~GeV (right) as simulated using FTFP\_BERT\_HP (red) or QGSP\_BERT\_HP (blue) physics lists of Geant4 version 10.3.}
\end{figure}

The distributions of calorimetric observables are shown in figures \ref{fig:obs_nhits_esum},  \ref{fig:obs_niso_ntrk},  \ref{fig:obs_ehit_rad_zcog}. For 10~GeV, the models agree within statistical uncertainties for all observables as expected. Above 10~GeV, a good agreement between models is observed for reconstructed energy and longitudinal centre of gravity. For other observables, the behaviour of QGSP\_BERT\_HP physics list is more smooth than that of FTFP\_BERT\_HP in the tails of distributions, while both physics lists give a well consistent predictions for the most probable values.  

The transverse segmentation with the step of 3~cm provides additional information about the radial energy density distribution within a shower. Having the transverse size of 72~cm, one can divide the calorimeter transverse plane into 12 rings around the shower centre of gravity, each ring being 3~cm thick, and calculate the so-called "ring" observables integrated over the longitudinal depth:  

\begin{itemize}
\item \textbf{Number of isolated hits in a ring}, $N_{j,\mathrm{iso}}$, accounts for hits, which isolation is defined as discussed above and for which the cell centre is within the $j$-th ring ($1 \leq j \leq 12$). 
\item \textbf{Energy sum in a ring}, $E_{j,\mathrm{ring}}$, is the sum of measured energy of all cells or parts of the cells, which belong to the $j$-th ring. For the cells shared by neighbour rings, the cell amplitude is also shared between the rings proportionally to the cell area attributed to the particular ring.
\end{itemize}

\section{Correlation between calorimetric observables and properties of secondaries}
\label{sec:corr}

\begin{figure}[htbp]
\centering 
\includegraphics[width=.45\textwidth]{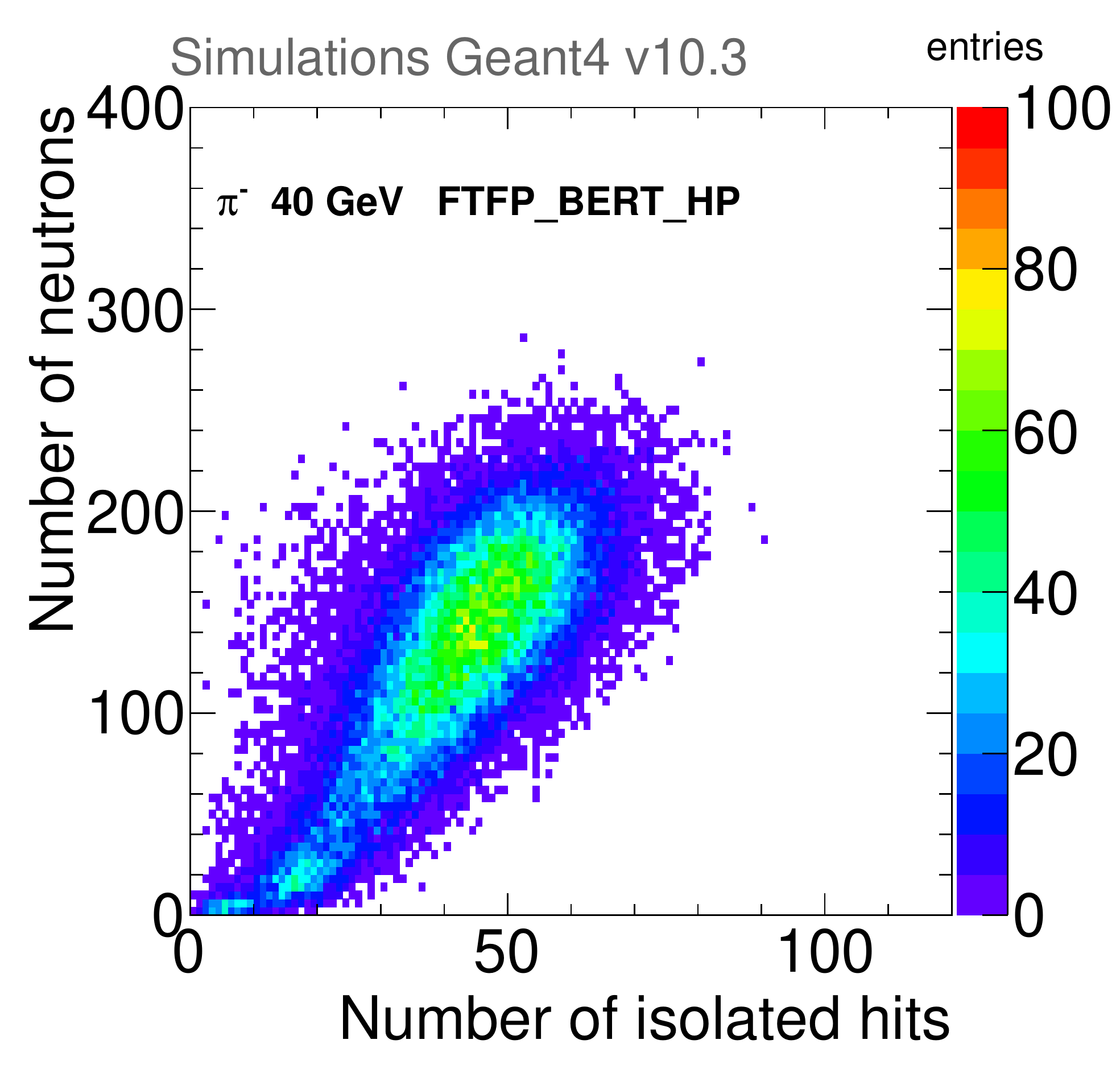}
\qquad
\includegraphics[width=.45\textwidth]{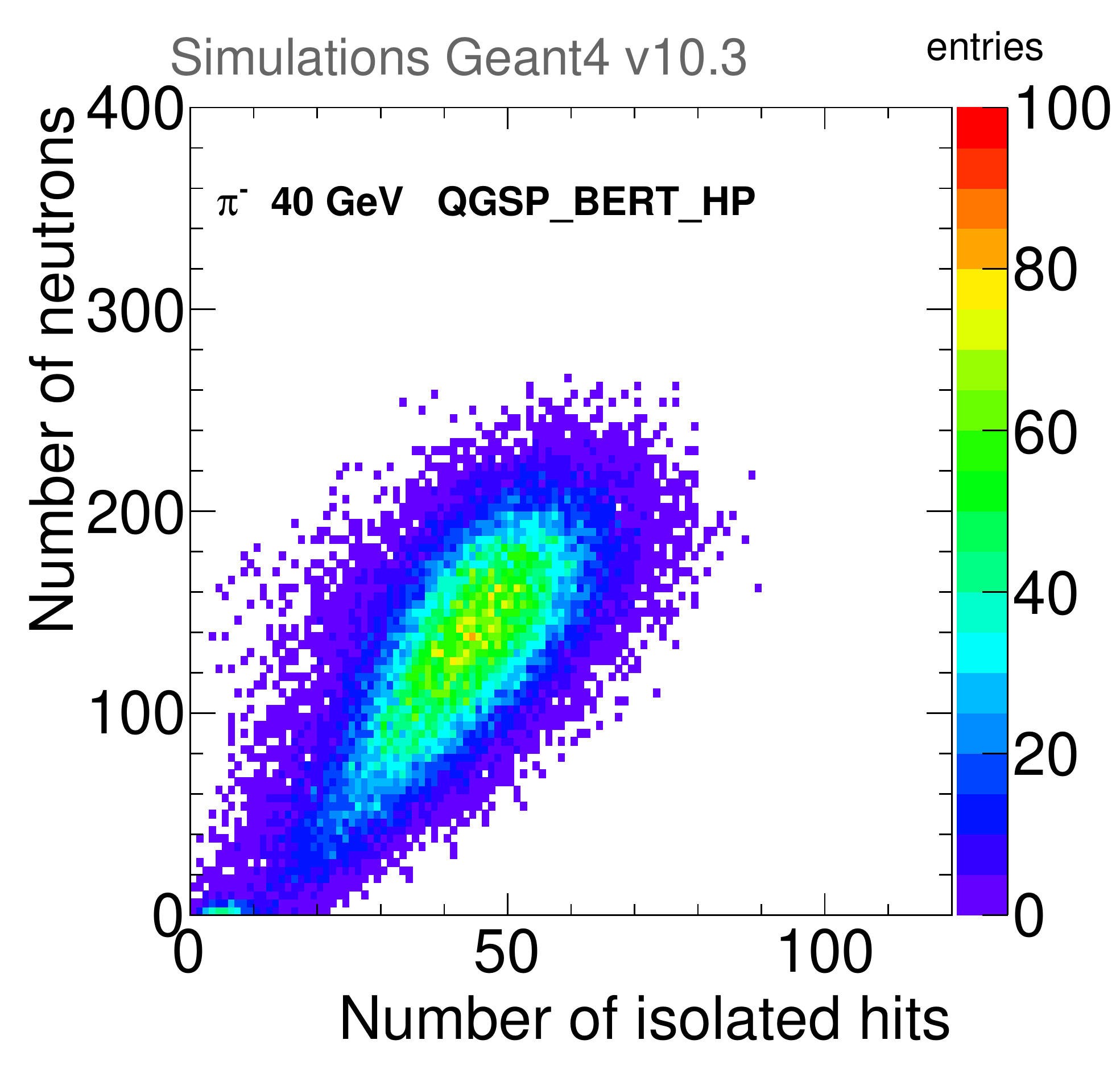}
\caption{\label{fig:corr_niso_nneu} Joint distribution of number of neutrons and number of isolated hits for hadronic showers initiated  by 40~GeV $\pi^{-}$  as simulated using FTFP\_BERT\_HP (left) or QGSP\_BERT\_HP (right) physics lists of Geant4 version 10.3.}
\end{figure}

\begin{figure}[htbp]
\centering 
\includegraphics[width=.45\textwidth]{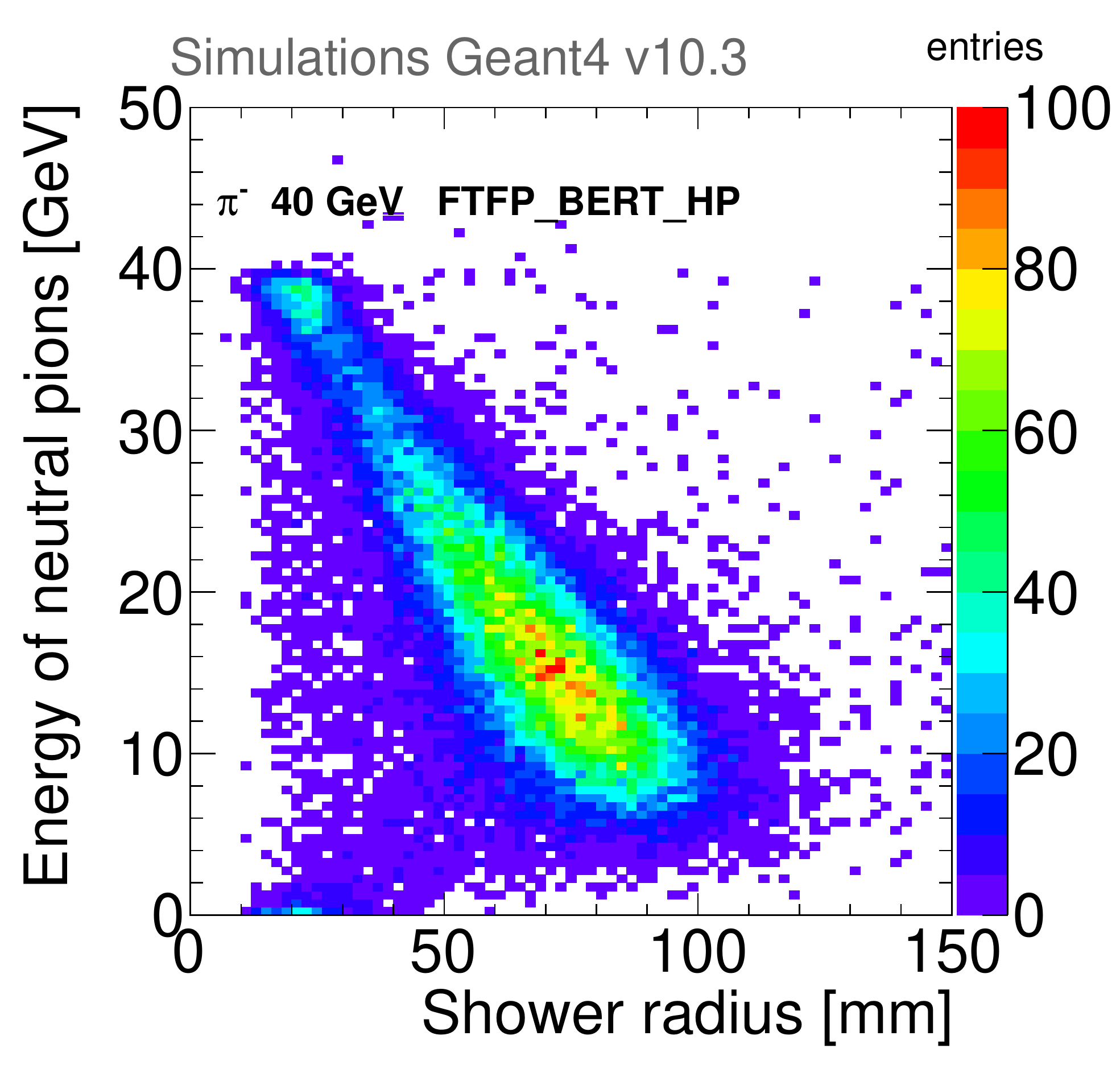}
\qquad
\includegraphics[width=.45\textwidth]{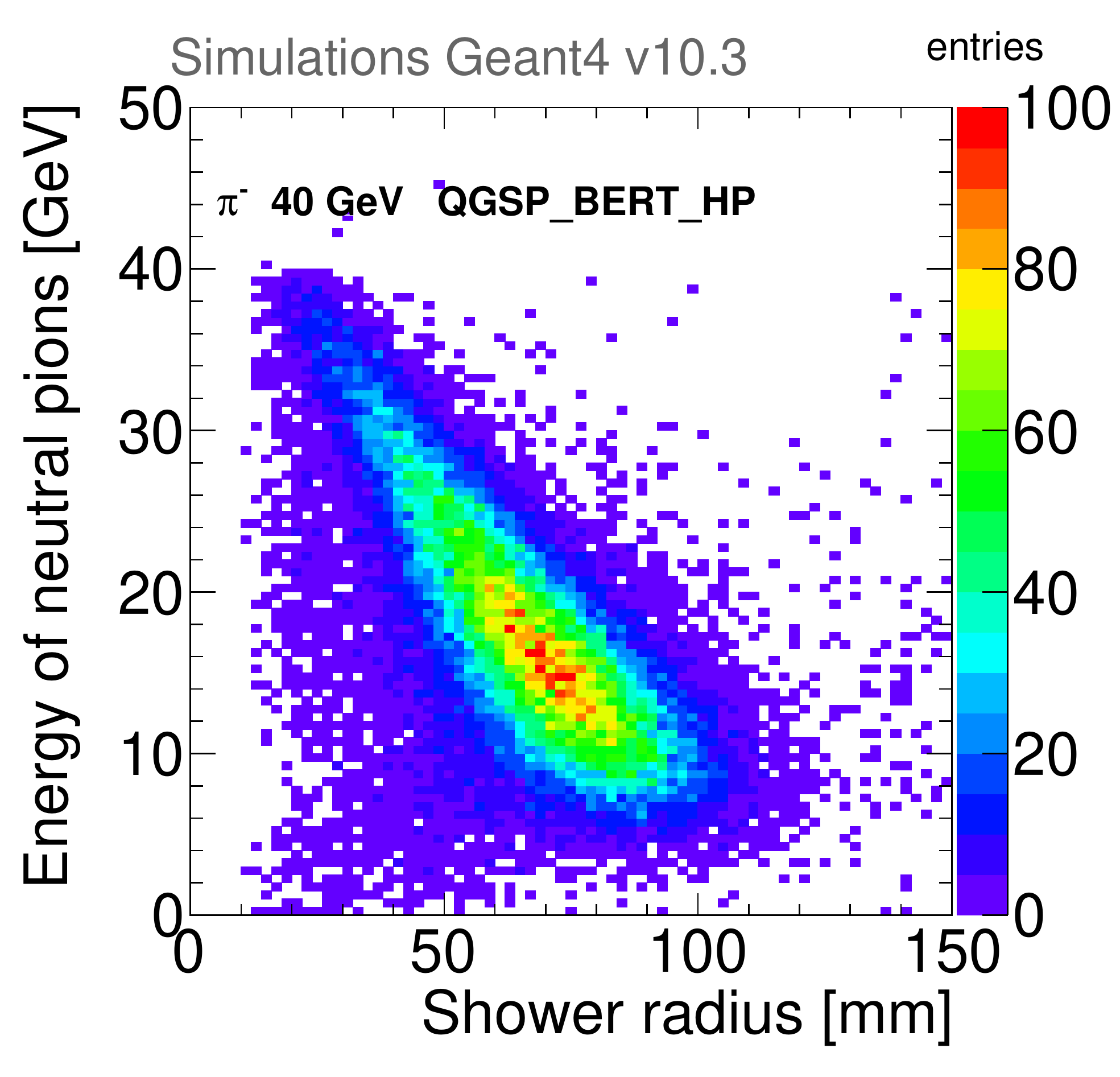}
\caption{\label{fig:corr_epi0_ering} Joint distribution of energy of neutral pions and shower radius for hadronic showers initiated by 40~GeV $\pi^{-}$  as simulated using FTFP\_BERT\_HP (left) or QGSP\_BERT\_HP (right) physics lists of Geant4 version 10.3.}
\end{figure}

Comparing the distributions of calorimetric observables and distributions of parameters of secondary particles produced in hadronic showers, one can suggest the relationship between some of them. Such relationships were investigated using joint distributions, where correlations and anti-correlations can be seen. Figure \ref{fig:corr_niso_nneu} represents the joint distribution of the number of isolated hits and number of neutrons in a shower and demonstrates the high correlation between calorimetric observable and intrinsic shower property. Figure \ref{fig:corr_epi0_ering} shows the joint distributions of the energy of neutral pions and the shower radius. It should be noted that in addition to the overall correlation observed, the peculiarities predicted by the FTFP\_BERT\_HP physics list are also correlated between two variables.   

\section{Regression model implementation in a deep neural network}
\label{sec:dnn}

The relationships observed between calorimetric and Monte-Carlo-truth variables open a possibility to use the calorimetric observables for the prediction of some intrinsic shower properties. At the same time, the complexity and nonlinearity of such relationships  as well as the quite large number of observables available make it difficult to implement analytic regression models. To solve this problem, a deep neural network was constructed and trained using the open-source library TensorFlow \cite{ref:tensorflow} within the python-based Keras framework.

\subsection{DNN structure, target and input variables}
\label{sec:dnn_struct}

The deep neural network structure applied for this regression task has a traditional feedforward architecture with the number of network layers, $L$, and the number of neurons per layer, $N_l$, where $l$ is the layer number.
The propagation of the signal through the network can be described in terms of the outputs, $y_{lj}$, of each neuron ($1 \leq j \leq N_l$) in each layer ($1 \leq l \leq L$):

\begin{equation}
\label{eq:dnn_func}
y_{lj} = \begin{cases}
x_{j}, \quad N_l = N_0  &\text{if $l = 1$;}\\
f_{lj}(s), \quad s = \sum\limits^{N_{l-1}}_{i=1}  w_{lij} \cdot y_{(l-1)i} + b_{lj}, & \text{if $1 < l < L$;}\\
\sum\limits^{N_{l-1}}_{i=1}  w_{lij} \cdot y_{(l-1)i} + b_{lj}, \quad N_l = 1, & \text{if $l = L$;}
\end{cases}
\end{equation}

\noindent where $x_{j}$ is the element of input vector $\vec x$ of size $N_0$, $f_{lj}$ is the activation function of $j$-th neuron in the layer $l$, $w_{lij}$ and $b_{lj}$ are the weights adjusted during a network training.

\begin{figure}[htbp]
\centering 
\includegraphics[width=.6\textwidth]{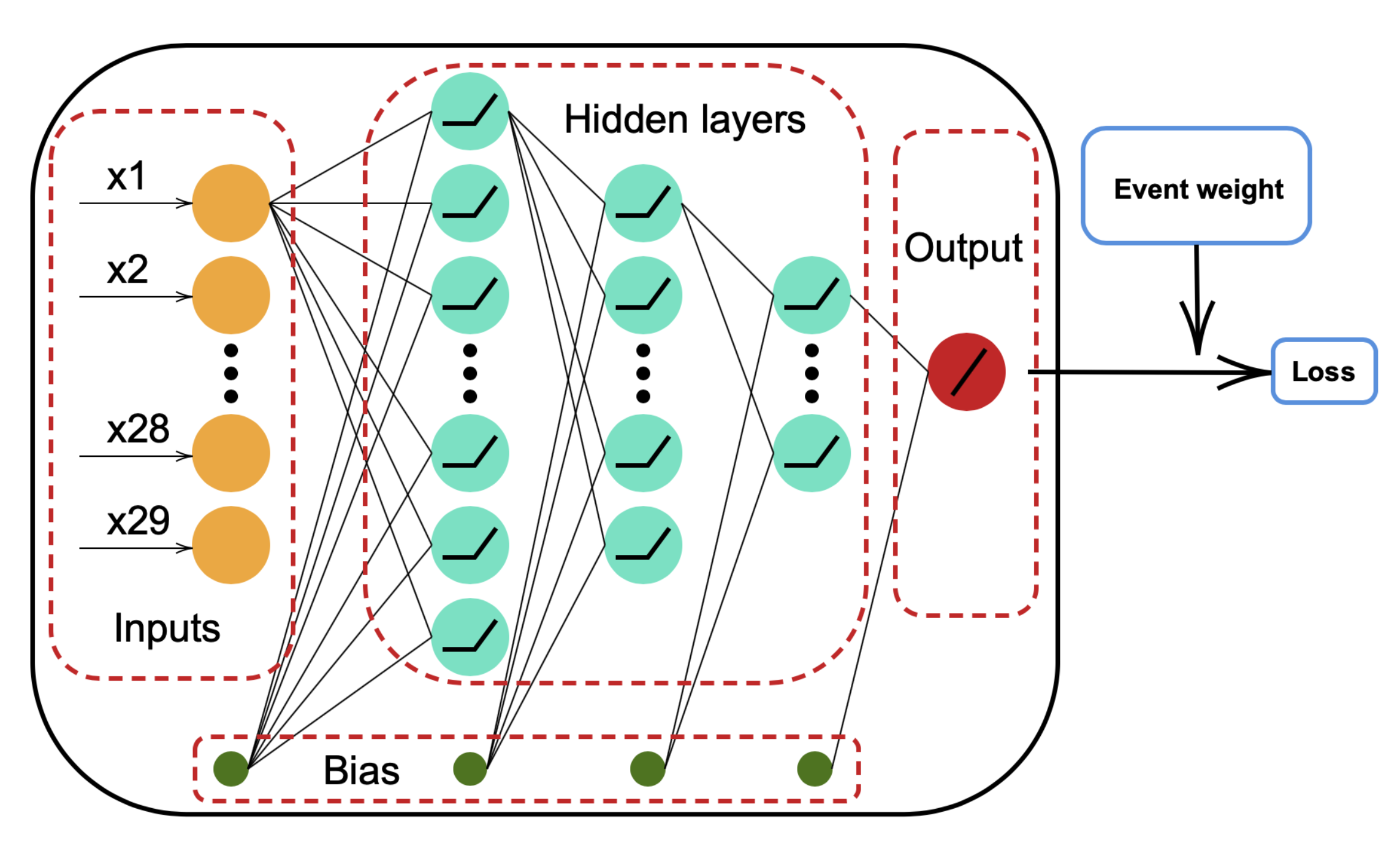}
\caption{\label{fig:dnn_arch} The structure of the deep neural network used in the study. For each layer, the links to the next layer are shown for the upper neuron only for better readability. See text for details.}
\end{figure}

In the DNN under study, the total number of layers $L = 5$: one input layer, three hidden layers and one output layer. The number of neurons in the input layer corresponds to the number of input variables: $N_1 = N_0 = $~29. The input variables are calorimetric observables described in section \ref{sec:calo_obs}.
The numbers of neurons in the hidden layers are $N_2 = $~128, $N_3 = $~64, $N_4 = $~32 with the ReLU activation function $f_{lj}(s) = \max(0,s)$ for each $j$-th neuron of layer $l$. The input and hidden layers also contain bias neurons, which introduce additive bias $b_{lj}$ to the argument of the activation function. The output layer consists of one neuron with linear activation function and returns the predicted value $Y^{\mathrm{pred}} = y_{L1}$ for a loss function. 
The DNN architecture is illustrated in figure \ref{fig:dnn_arch}.

\subsection{DNN training and optimisation}
\label{sec:dnn_opt}

The supervised learning means the minimisation of the loss function, $F_{\mathrm{loss}}$, which characterises the difference between the predicted and true target values, by adjusting the weights $w_{lij}$ and $b_{lj}$. The total number of adjusted DNN parameters in the configuration described in section \ref{sec:dnn_struct} is about 14000. For the DNN training the Huber loss function was used: 	

\begin{equation}
\label{eq:dnn_loss}
F_{\mathrm{loss}} = \frac{1}{K} \cdot \sum\limits_{k=1}^{K} V_{k} \cdot \Delta Y_k, \quad
\Delta Y_k = \begin{cases}
0.5 \cdot (Y^{\mathrm{pred}}_{k} - Y^{\mathrm{true}}_{k})^{2}, & \quad \text{if} \quad |Y^{\mathrm{pred}}_{k} - Y^{\mathrm{true}}_{k}| \leq 1;\\
|Y^{\mathrm{pred}}_{k} - Y^{\mathrm{true}}_{k}| - 0.5, & \quad \text{if} \quad |Y^{\mathrm{pred}}_{k} - Y^{\mathrm{true}}_{k}| > 1;
\end{cases}
\end{equation}

\noindent where $K$ is the number of events in the training (or validation) sample and $V_k$ is the event weight. 

\subsubsection{Density-based reweighting}
\label{sec:dnn_reweight}

The event weights can take into account the non-uniformity of the training sample in the range studied (density-based weighting \cite{ref:dens-weight}) as well as the uncertainties or the stochastic nature of the true value of the parameter under investigation. The reweighting procedure is implemented during the training to improve the uniformly of a target variable and to avoid a possible shift of predictions to the most probable value.

The generalised representation of the density-based weighting
can be introduced as an inverse probability density distribution of the variable under study, $PDF_{\mathrm{hist}}$, extracted form the histogram with the relatively fine binning. The smoothed histogram, $PDF_{\sigma}(z_{i})$, is calculated using bin-by-bin convolution of the histogram $PDF_{\mathrm{hist}}(z_{i})$ with the Gaussian function, $G(z_{i}, z,\sigma)$, where $z_{i}$ is the bin center of the $i$-th bin and the width $\sigma$ defines the degree of smoothing. 

\begin{equation}
\label{eq:dnn_pdfconv}
PDF_{\sigma}(z_{i}) = (PDF_{\mathrm{hist}}(z_{i}) \ast G(z_{i},z,\sigma)) 
\end{equation}

For the particular level of smoothing, the weight of $k$-th event is calculated as

\begin{equation}
\label{eq:dnn_pdfweight}
V_{k}(\sigma, Y_k^{\mathrm{true}}) = a \cdot PDF^{-1}_{\sigma}(z_{i})
\end{equation}

\noindent where $Y_k^{\mathrm{true}}$ belongs to the bin with the bin centre $z_{i}$ and $a$ is the normalisation constant, which can help to avoid too low weight values in case of a large sample size and coarse binning. 

\subsubsection{DNN hyperparameters}
\label{sec:dnn_hyper}

Several DNN hyperparameters --- optimiser type, learning rate, batch size --- were investigated without dedicated optimisation. The detailed explanation of these hyperparameters meaning in the TensorFlow library can be found in ref.~\cite{ref:tensorflow}. The tested optimisers ADAM and NADAM have demonstrated similar performance. From the wide range of learning rates studied (from 10 to 10$^{-9}$), the values of 10$^{-6}$ and 10$^{-7}$ helped to achieve better performance for this task. Different batch sizes were also tested in the range from 1 to 256 events per batch. With increasing the batch size a slight decrease of result stability was observed. The batch size equal 8 has been chosen as a compromise between the convergence stability and the speed of calculation. 

After the event selection as described in section \ref{sec:conditions}, the remaining number of events was about 96000 for the FTFP\_BERT\_HP physics list and about 106000 for the QGSP\_BERT\_HP physics list. For equal DNN training conditions and further direct comparisons, 95000 events per each sample were used and divided into three subsamples in the following proportion: 60000 events for training, 15000 for validation and 20000 for test. The important hyperparameter is the number of epochs for training that is typically a compromise between  the convergence and overtraining trends. In the current study, for the given number of events, the convergence without overtraining was achieved between 15 and 50 training epochs.

\subsubsection{Feature importance investigation}	
\label{sec:dnn_importance}

In this study, the maximum number of calorimetric observables used as DNN inputs is 29 (as described in section \ref{sec:calo_obs}). The feature importance was investigated using the permutation technique from the scikit-learn package. This inspection method looks at the changes in the model performance when one of the features is randomly shuffled and its relationship with the target is artificially broken. The changes in performance are expressed in the importance score that can be calculated using different metrics. For both targets, three metrics were tried (coefficient of determination, mean squared error and explained variance) and resulted in the same order of feature importance for the particular target. For the prediction of the number of neutrons, the most important observables are: the energy in the most central ring around the shower axis, the number of isolated hits in a shower and the shower radius. For the prediction of the energy of neutral pions, the observables with the highest importance are: the energy in the most central and next-to-central rings around the shower axis and the shower radius.

The model was trained using different subsets of features with the highest importance defined by the inspection technique mentioned above. Though some of the features have very low importance score compared to the most important ones, it was found that the best performance is achieved for the full set of calorimetric observables as inputs. This result can be explained by the non-negligible correlations between several input features, which might lead to shadowing of the real importance in the inspection technique.

\section{Results}
\label{sec:res}

The deep neural network with the architecture described in section \ref{sec:dnn} was trained using the supervised learning on the simulated samples of hadronic showers induced by single pions with initial energy of 40~GeV in the model of the highly granular calorimeter (see section \ref{sec:calo_ahcal}). Two target variables are investigated: number of secondary neutrons and energy of neutral pions produced within a hadronic shower. For each target variable, two DNN-based regression models were trained on the samples generated using the QGSP\_BERT\_HP and FTFP\_BERT\_HP physics lists from Geant4 package version 10.3. 

In this study, the estimate of DNN performance was based on the comparison of three momenta (mean, median and standard deviation) of the true and predicted distributions for the test subsample of the same physics list, on which the model was trained. For all the examples shown below, the agreement between true and predicted momenta is within few percent and for the majority of cases, the agreement in asymmetry is within 10\%.
The crosscheck was also performed by applying each of the trained regression models to the test subsample from another physics list. 

\subsection{Prediction of the number of neutrons in a hadronic shower}
\label{sec:res_neutron}

\begin{figure}[htbp]
\centering 
\includegraphics[width=.45\textwidth]{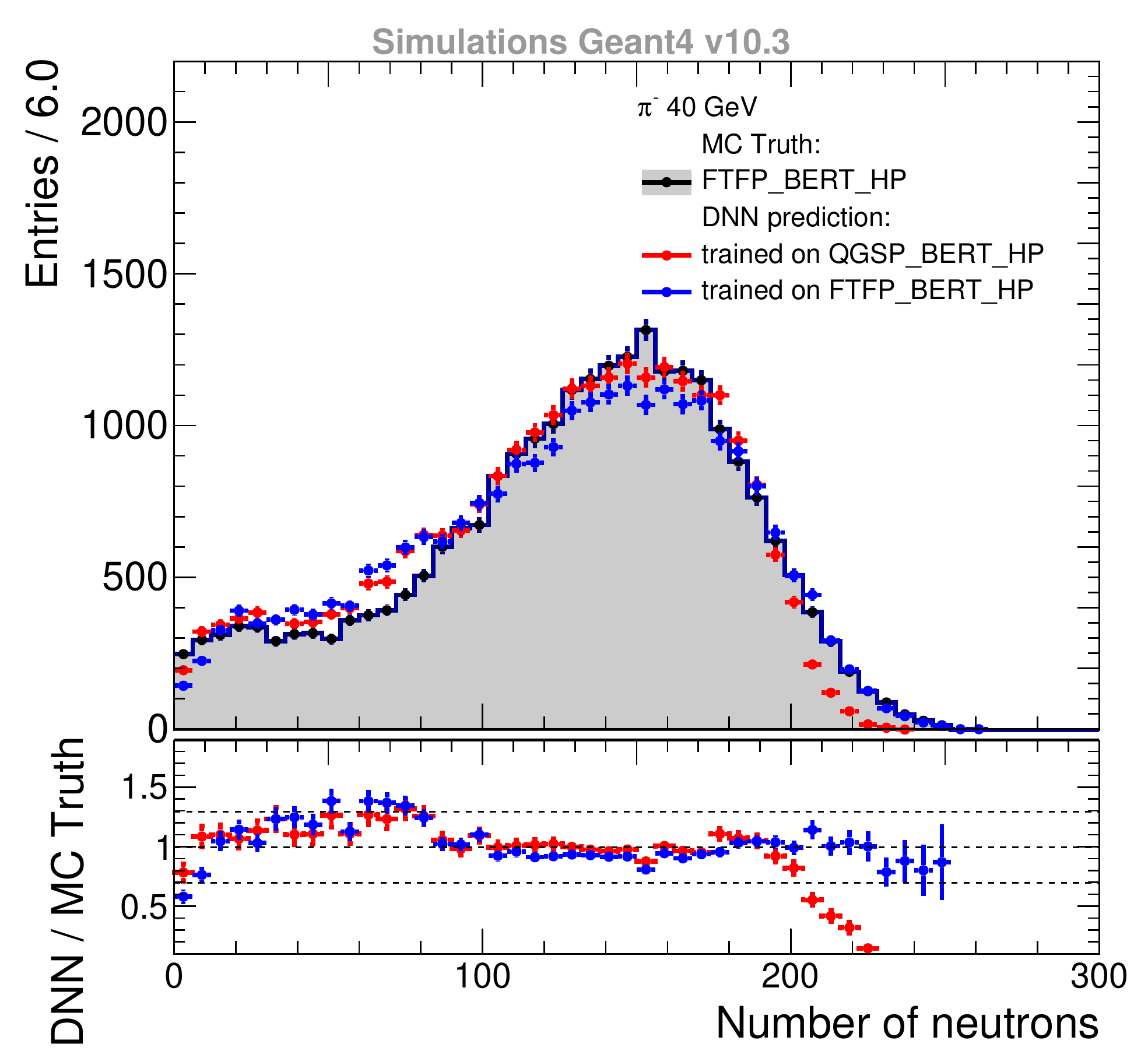}
\qquad
\includegraphics[width=.45\textwidth]{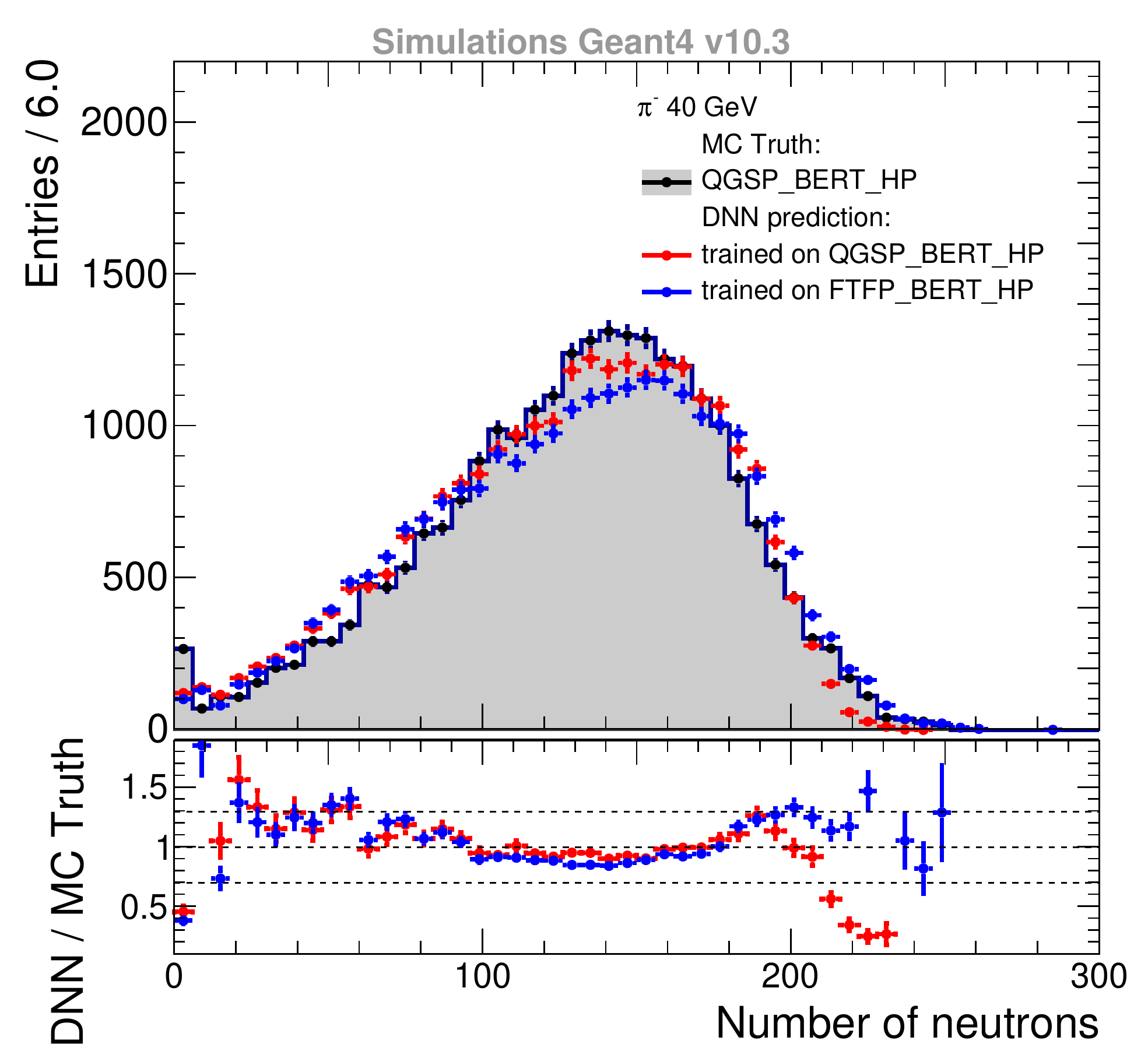}
\caption{\label{fig:dnnNnDist} Distribution of the number of neutrons in a hadronic shower initiated by 40~GeV $\pi^{-}$ generated by FTFP\_BERT\_HP (grey, left) or QGSP\_BERT\_HP (grey, right) physics lists and predicted by the DNN regression models trained on the samples produced using QGSP\_BERT\_HP (red) or FTFP\_BERT\_HP (blue) physics list. The bottom plots contain ratios of predicted to true distributions.}
\end{figure}

The distributions of the generated and predicted number of neutrons in a hadronic shower induced by a 40~GeV pion in the AHCAL are presented in figure \ref{fig:dnnNnDist} for two physics lists. Two predictions are shown for both cases: from the models trained on the QGSP\_BERT\_HP and on the FTFP\_BERT\_HP physics list. The model trained on the FTFP\_BERT\_HP physics list gives better prediction for the large number of neutrons, while the model trained on the QGSP\_BERT\_HP physics list provides better coincidence around the most probable value. The discrepancy at the tails can be explained by the choice of performance criteria, which are based on momenta comparison and not on bin-by-bin agreement of the distributions.

\begin{figure}[htbp]
\centering 
\includegraphics[width=.45\textwidth]{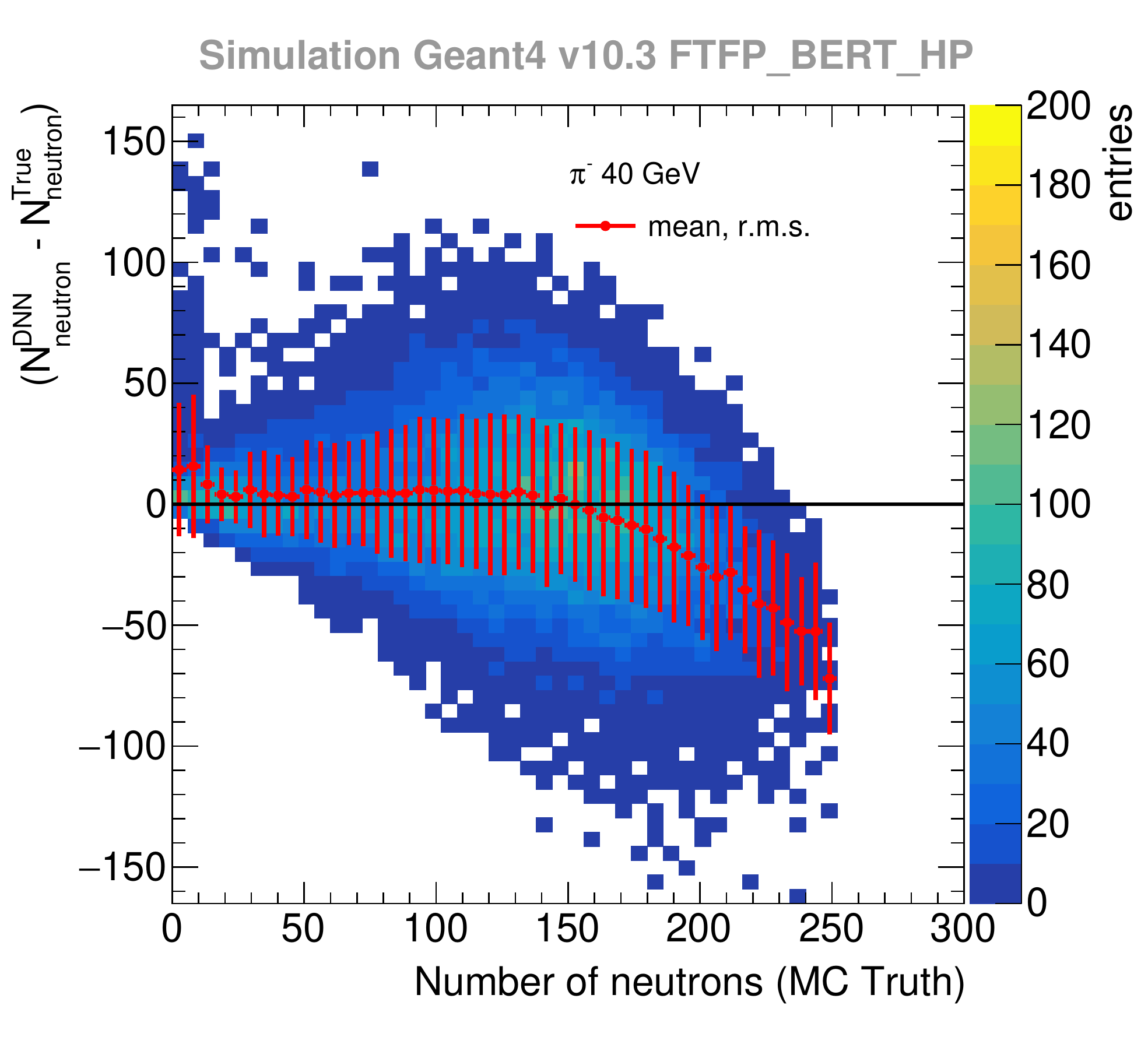}
\qquad
\includegraphics[width=.45\textwidth]{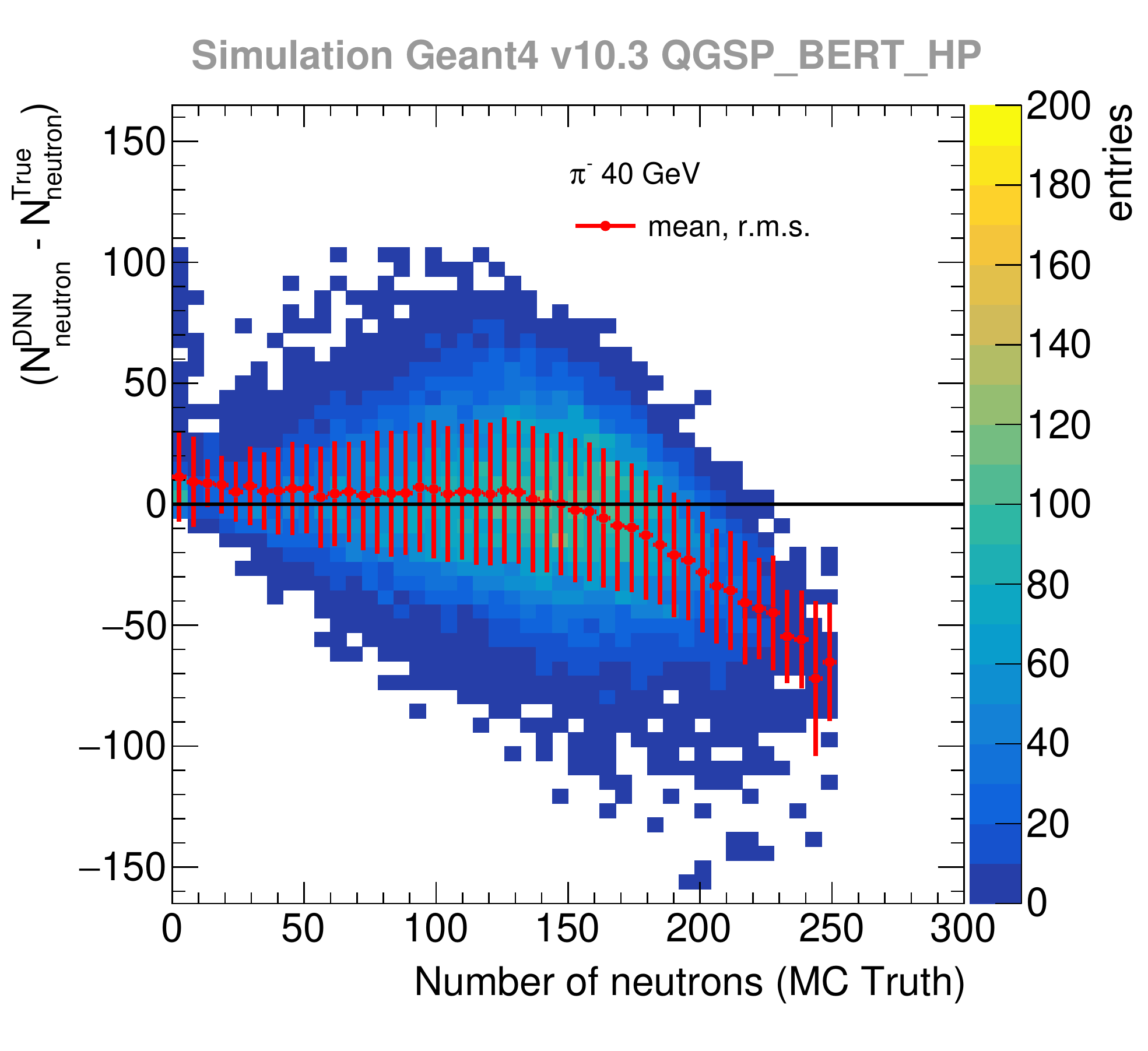}
\caption{\label{fig:dnnNn2D} Joint distribution of the residuals between generated and predicted number of neutrons in a hadronic shower initiated by 40~GeV $\pi^{-}$ and generated values. The red points and error bars correspond to the means and standard deviations of residuals in the bins of generated values. 
The DNN regression models used for predictions are trained on the same physics lists (left: FTFP\_BERT\_HP, right: QGSP\_BERT\_HP).
}
\end{figure}

The residuals between generated ($N^{\mathrm{true}}_{\mathrm{neutron}}$) and predicted ($N^{\mathrm{pred}}_{\mathrm{neutron}}$) number of neutrons versus generated ones are shown in figure \ref{fig:dnnNn2D}. While the first, second and third momenta of the true distributions are quite well reproduced, the uncertainty of event-by-event predictions is $\sim$10--15\% around the most probable value. The largest deviations are observed for large number of neutrons, where both trained models underestimate the true values.

\subsection{Prediction of the energy sum of $\pi^{0}$s in a hadronic shower}
\label{sec:res_epi0}

\begin{figure}[htbp]
\centering 
\includegraphics[width=.45\textwidth]{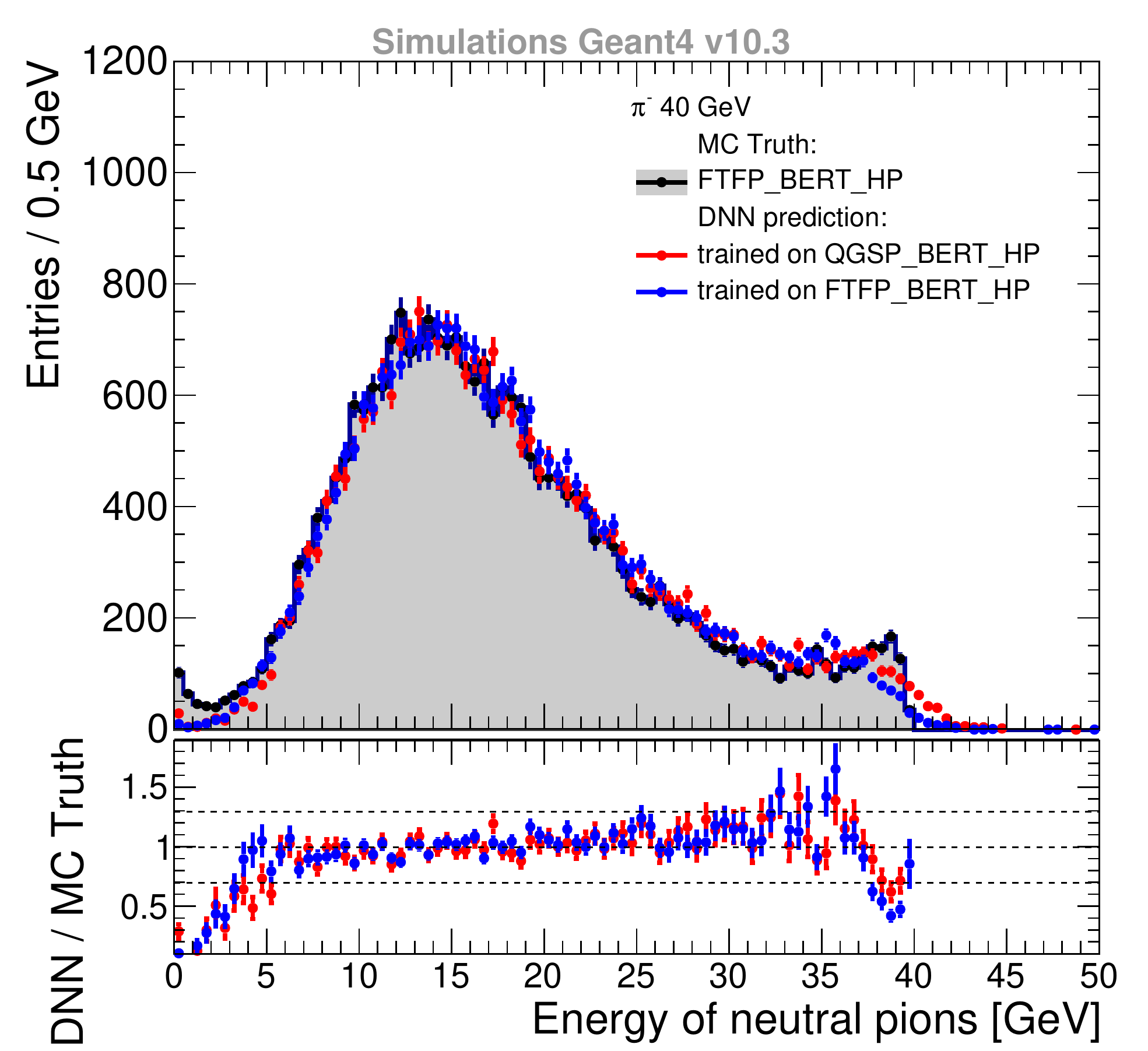}
\qquad
\includegraphics[width=.45\textwidth]{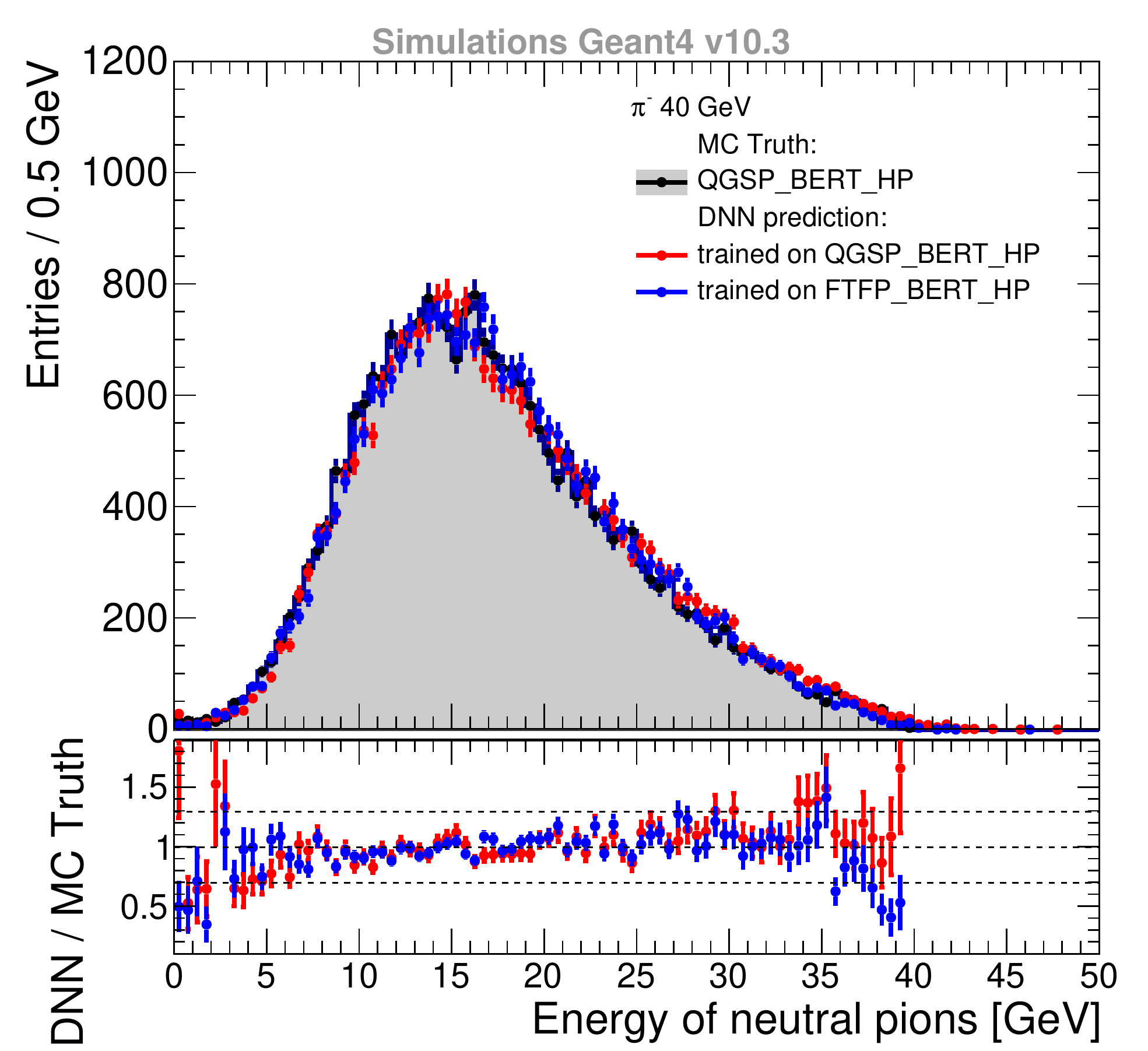}
\caption{\label{fig:dnnEpi0Dist} Distribution of the energy of neutral pions in a hadronic shower initiated by 40~GeV $\pi^{-}$ generated by FTFP\_BERT\_HP (grey, left) or QGSP\_BERT\_HP (grey, right) physics lists and predicted by the DNN regression models trained on the samples produced using QGSP\_BERT\_HP (red) or FTFP\_BERT\_HP (blue) physics list. The bottom plots contain ratios of predicted to true distributions.}
\end{figure}

The distributions of the energy of neutral pions shown in figure~\ref{fig:dnnEpi0Dist} are well reproduced by both DNN regression models. The largest deviation of predictions from the true distribution is observed for the FTFP\_BERT\_HP physics list (figure~\ref{fig:dnnEpi0Dist}, left) in the regions of non-smooth behaviour, below 3~GeV and above 37~GeV.  The discrepancy in the boundary regions indicates the limitations of either the model architecture or the discriminative power of the calorimetric observables used. The architecture optimisation requires the larger sample size to allow increase of the number of DNN parameters, while the set of available input variables is constrained by the calorimeter design.    

\begin{figure}[htbp]
\centering 
\includegraphics[width=.45\textwidth]{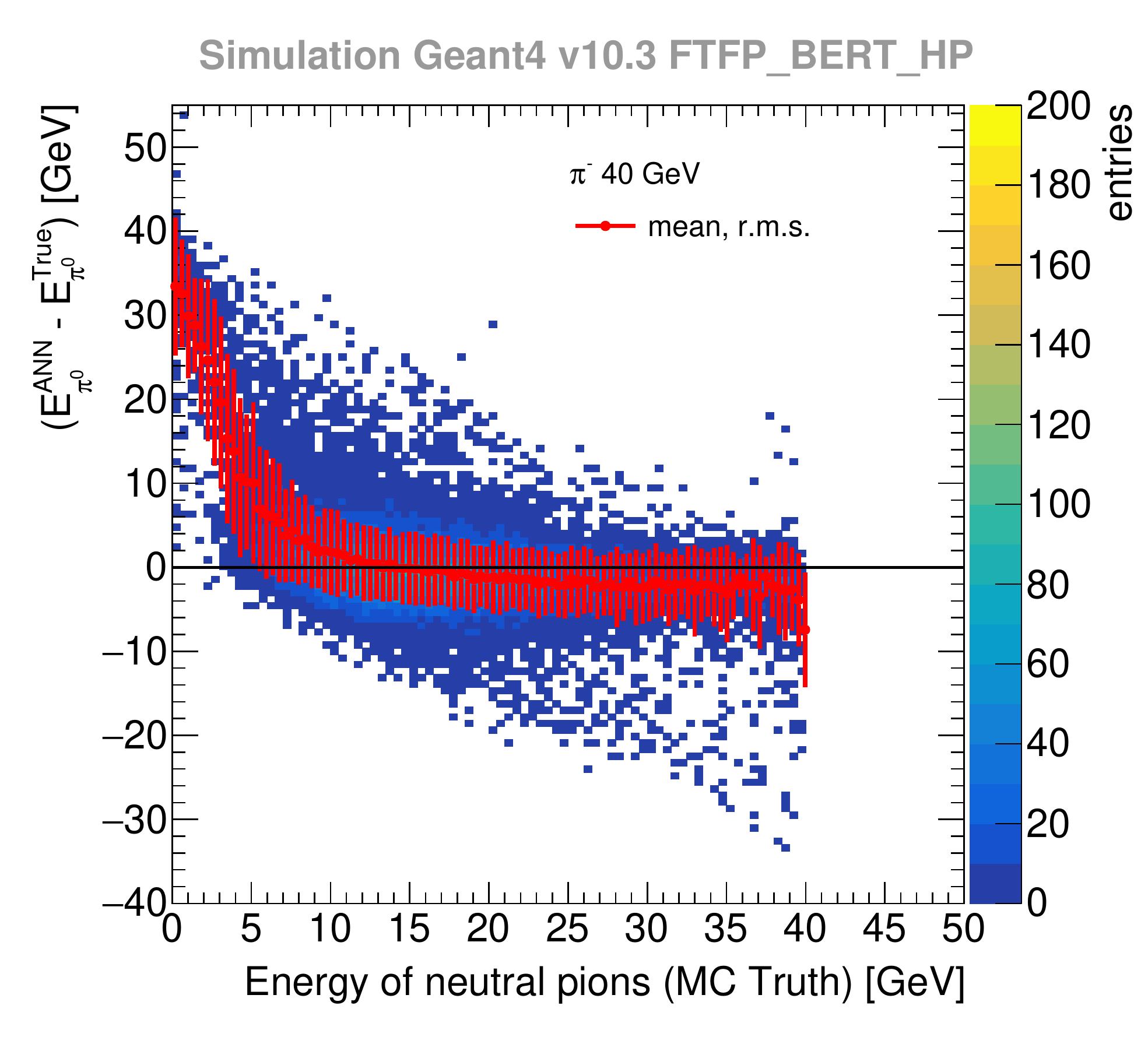}
\qquad
\includegraphics[width=.45\textwidth]{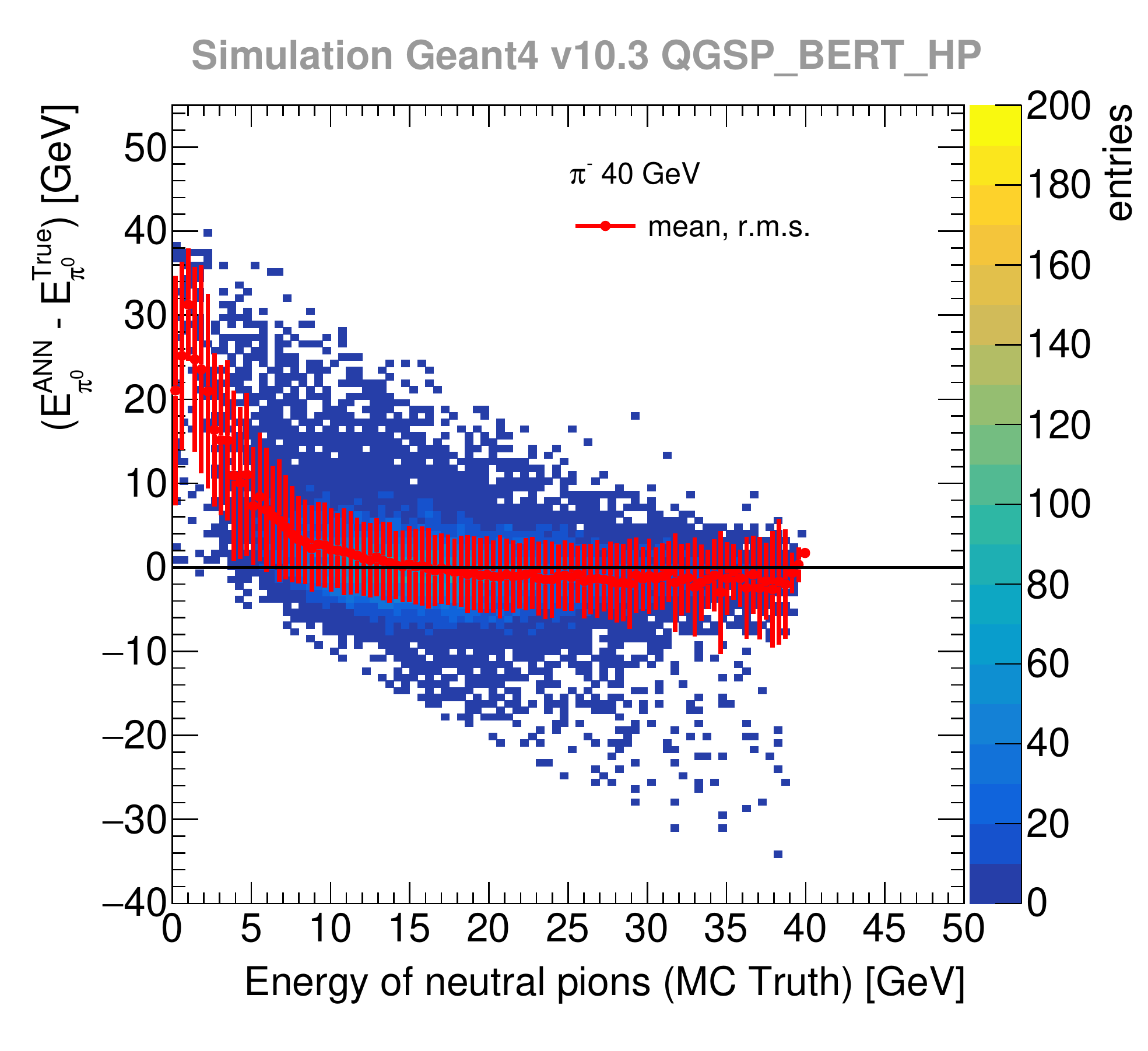}
\caption{\label{fig:dnnEpi02D} Joint distribution of the residuals between generated and predicted energy of neutral pions in a hadronic shower initiated by 40~GeV $\pi^{-}$ and generated values. The red points and error bars correspond to the means and standard deviations of residuals in the bins of generated values. 
The DNN regression models used for predictions are trained on the same physics lists (left: FTFP\_BERT\_HP, right: QGSP\_BERT\_HP).
}
\end{figure}

The event-by-event comparison of the predicted and true energy of neutral pions is illustrated in figure~\ref{fig:dnnEpi02D}. The largest difference between predicted and true values is observed for the small values of true neutral pion energy (below 3~GeV). The fraction of such events with unreliable predictions is $\sim$0.4\% for the QGSP\_BERT\_HP physics list and $\sim$1.4\% for the FTFP\_BERT\_HP physics list.

\section{Conclusion}
\label{sec:conclusion}
The new approach proposed in this work exploits the relationships between calorimetric observables and the properties of secondaries, which are observed for the simulated hadronic showers.  
To test the approach and estimate its performance, showers initiated by 40~GeV negative pions are used, which are simulated in the model of the highly granular analog hadron calorimeter with two physics lists of Geant4 package version 10.3. Several examples of deep neural networks were trained with a supervised learning to predict the number of secondary neutrons or the energy of neutral pions within a hadronic shower. The relatively simple DNN-based regression models show good performance and predict the mean, median and standard deviation of the target distributions at few percent level. 

A direct comparison of data and simulations is not straightforward due to model dependence on the physics list used for training. At the same time, the application of the trained DNN model to test beam data and comparison of predicted distributions for number of neutrons and energy of neutral pions will provide additional information for the validation of simulations.
Another application, which is currently under investigation, was inspired by the reasonable performance achieved by the trained DNN models in the event-by-event prediction of the energy of neutral pions. This energy is closely related to the electromagnetic fraction in hadronic showers and can be used to correct measured hadron energy and improve the energy resolution. 

\acknowledgments

The authors are very grateful to the CALICE colleagues for the availability to use centrally produced raw simulated samples as well as for the very fruitful discussions and comments.


\begin{thebibliography}{99}

\bibitem{ref:Geant4-2003}
S.~Agostinelli \emph{et al.}, 
\emph{Geant4--a simulation toolkit},
\emph{Nucl. Instrum. Meth.} \textbf{A506} (2003) 250.

\bibitem{ref:Geant4-2016}
J.~Allison \emph{et al.}, 
\emph{Recent developments in Geant4},
\emph{Nucl. Instrum. Meth.} \textbf{A835} (2016) 186.

\bibitem{ref:PFA-2009ilc}
M.~A.~Thomson, 
\emph{Particle Flow Calorimetry and the PandoraPFA Algorithm}, 
\emph{Nucl. Instrum. Meth.}  \textbf{A611} (2009) 25.

\bibitem{ref:SiECAL-2008}
J.~Repond  \emph{et al.}, 
\emph{Design and electronics commissioning of the physics prototype of a Si-W electromagnetic calorimeter for the International Linear Collider}, 
\emph{JINST} \textbf{3} (2008) P08001.

\bibitem{ref:AHCAL-2010cc}
C.~Adloff \emph{et al.}, 
\emph{Construction and commissioning of the CALICE analog hadron calorimeter prototype}, 
\emph{JINST} \textbf{5} (2010) P05004.

\bibitem{ref:SDHCAL-2015}
G.~Baulieu \emph{et al.},
\emph{Construction and commissioning of a technological prototype of a high-granularity semi-digital hadronic calorimeter},
\emph{JINST} \textbf{10} (2015) P010039.

\bibitem{ref:ScECAL-2018}
J.~Repond \emph{et al.},
\emph{Construction and Response of a Highly Granular Scintillator-based Electromagnetic Calorimeter}, 
\emph{Nucl. Instrum. Meth.} \textbf{A887} (2018) 150. 

\bibitem{ref:AHCAL-2013tracks}
C.~Adloff \emph{et al.}, 
\emph{Track segments in hadronic showers in a highly granular scintillator-steel hadron calorimeter}, 
\emph{JINST} \textbf{8} (2013) P09001.

\bibitem{ref:SDHCAL-2017tracks}
Z.~Deng \emph{et al.},
\emph{Tracking within Hadronic Showers in the CALICE SDHCAL prototype using a Hough Transform Technique},
\emph{JINST} \textbf{12} (2017) P05009.

\bibitem{ref:AHCAL-2013valid}
C.~Adloff \emph{et al.}, 
\emph{Validation of GEANT4 Monte Carlo Models with a Highly Granular Scintillator-Steel Hadron Calorimeter}, 
\emph{JINST} \textbf{8} (2013) P07005.

\bibitem{ref:SiWECAL-2015}
B.~Bilki \emph{et al.},
\emph{Testing hadronic interaction models using a highly granular silicon-tungsten calorimeter},
\emph{Nucl. Instr. Meth.} \textbf{A794} (2015) 240.

\bibitem{ref:AHCAL-2016dec} 
G.~Eigen \emph{et al.},
\emph{Hadron shower decomposition in the highly granular CALICE analogue hadron calorimeter},
\emph{JINST} \textbf{11} (2016) P06013. 

\bibitem{ref:WAHCAL-2015}
M.~Chefdeville \emph{et al.}, 
\emph{Shower development of particles with momenta from 15 GeV to 150 GeV in the CALICE scintillator-tungsten hadronic calorimeter}, 
\emph{JINST} \textbf{10} (2015) P12006.

\bibitem{ref:DHCAL-2019}
M.~Chefdeville \emph{et al.},
\emph{Analysis of Testbeam Data of the Highly Granular RPC-Steel CALICE Digital Hadron Calorimeter and Validation of Geant4 Monte Carlo Models}, 
\emph{Nucl. Instr. Meth.} \textbf{A939} (2019) 89.

\bibitem{ref:procVCI-2019}
R.~P\"oschl,
\emph{Recent results of the technological prototypes of the CALICE highly granular calorimeters},
\emph{Nucl. Instrum. Meth.} \textbf{A958} (2020) 162234.

\bibitem{ref:ML-2002}
J.~Kieseler,
\emph{Object condensation: one-stage grid-free multi-object reconstruction in physics detectors, graph, and image data},
\emph{Eur. Phys. J.} \textbf{C80} (2020) 886. 

\bibitem{ref:ML-2018review}
D.~Guest, K.~Cranmer and D.~Whiteson,
\emph{Deep Learning and Its Application to LHC Physics},
\emph{Annu. Rev. Nucl. Part. Sci.} \textbf{68} (2018) 1.

\bibitem{ref:ML-2021gnn}
J.~Pata, \emph{et al.}, 
\emph{MLPF: efficient machine-learned particle-flow reconstruction using graph neural networks},
\emph{Eur. Phys. J.} \textbf{C81} (2021) 381. 

\bibitem{ref:ML-2020calo}
D.~Belayneh \emph{et al.},
\emph{Calorimetry with Deep Learning: Particle Simulation and Reconstruction for Collider Physics},
\emph{Eur. Phys. J.} \textbf{C80} (2020) 688. 

\bibitem{ref:ML-2020bjet}
A.~M.~Sirunyan, \emph{et al.}, 
\emph{A Deep Neural Network for Simultaneous Estimation of b Jet Energy and Resolution}, 
\emph{Comput. Softw. Big Sci.} \textbf{4} (2020) 10. 

\bibitem{ref:ML-2022kinem}
M.~Arratia \emph{et al.},
\emph{Reconstructing the Kinematics of Deep Inelastic Scattering with Deep Learning},
\emph{Nucl. Instrum. Meth.} \textbf{A1025} (2022) 166164 

\bibitem{ref:AHCALtest}
O.~Pinto (for the CALICE collaboration),
\emph{Operation and Calibration of a Highly Granular Hadron Calorimeter with SiPM-on-Tile Read-out},
Proceedings, 2019 IEEE Nuclear Science Symposium (NSS) and Medical Imaging Conference (MIC) (NSS/MIC 2019) : Manchester, United Kingdom, October 26- November 02, 2019. 

\bibitem{ref:G4-2016ftfp}
J.~Allison \emph{et al.}, 
\emph{Recent developments in Geant4},
\emph{Nucl. Instr. Meth.} \textbf{A835} (2016) 186.

\bibitem{ref:G4-2009qgsp}
J.~Apostolakis  \emph{et al.}, 
\emph{Geometry and physics of the Geant4 toolkit for high and medium energy applications},
\emph{Radiation Physics and Chemistry} \textbf{78} (2009) 859. 

\bibitem{ref:tensorflow}
M.~Abadi \emph{et al.}, 
\emph{Tensorflow: A system for large-scale machine learning}, 
\href{https://doi.org/10.5281/zenodo.5043456}{doi.org/10.5281/zenodo.5043456}, 2016.

\bibitem{ref:dens-weight}
M.~Steininger \emph{et al.}, 
\emph{Density-based weighting for imbalanced regression},
\emph{Mach Learn} \textbf{110} (2021) 2187. 


\end{thebibliography}
\end{document}